\documentclass[twocolumn, superscriptaddress, nofootinbib]{revtex4-1}

\usepackage[english]{babel}
\usepackage[pdftex]{graphicx}
\usepackage{color}
\usepackage{ulem}

\usepackage[sort&compress]{natbib}
\usepackage[colorlinks=true,allcolors=blue]{hyperref}
\usepackage{chngcntr}
\usepackage{url}
\usepackage{amsmath}
\usepackage{amssymb}
\usepackage{mathtools}
\usepackage{braket}

\definecolor{bluenew}{rgb}{0.07, 0.53, 0.37}
\renewcommand{\vec}[1]{\ensuremath{\boldsymbol{#1}}}

\begin{document}
\title{Valley-dependent time evolution of coherent electron states\\ 
in tilted anisotropic Dirac materials}

\author{Yonatan Betancur-Ocampo}
\email{ybetancur@icf.unam.mx}
\affiliation{Instituto de Ciencias F\'isicas, Universidad Nacional Aut\'onoma de M\'exico, Cuernavaca, M\'exico}

\author{Erik D\'iaz-Bautista}
\email{ediazba@ipn.mx}
\affiliation{Unidad Profesional Interdisciplinaria de Ingenier\'ia Campus Hidalgo, Instituto Polit\'ecnico Nacional, Pachuca, M\'exico}

\author{Thomas Stegmann}
\email{stegmann@icf.unam.mx}
\affiliation{Instituto de Ciencias F\'isicas, Universidad Nacional Aut\'onoma de M\'exico, Cuernavaca, M\'exico}

\begin{abstract}
The effect of the Dirac cone tilt of anisotropic two-dimensional materials on the time evolution of coherent electron states in the presence of electric and magnetic fields is studied. We propose a canonical transformation that maps the anisotropic Dirac-Weyl Hamiltonian with tilted Dirac cones to an effective and isotropic Dirac Hamiltonian under these fields. In this way, the well-known Landau-level spectra and wave functions allow calculating the Wigner matrix representation of Landau and coherent states. We found a valley dependency in the behavior of the Wigner function for both Landau and coherent electron states. The time evolution shows that the interplay of the Dirac cone tilt and the electric field keeps the uncertainties of both position and momentum in one valley significantly lower than in the other valley. The increment of quantum noise correlates with the emergence of negative values in the Wigner function. These results may help us to understand the generation of coherent electron states under the interaction with electromagnetic fields. The reported valley-dependent signatures in the Wigner function of materials with tilted Dirac cones may be revealed by quantum tomography experiments, even in the absence of electric fields.
\end{abstract}

\maketitle

\section{Introduction}
Most of the two-dimensional materials discovered so far, such as borophene, strained graphene, Weyl semimetals, and the organic conductor $\alpha$-(BEDT-TTF)$_2$I$_3$ present anisotropy and tilted Dirac cones in their low energy band structure \cite{Zhou2014, Mannix2015, Feng2016, Zabolotskiy2016, LopezBezanilla2016, Li2018, Wang2019, Zhang2018, Zhang2019, Tajima2009, Goerbig2009, Armitage2018, Ferreira2021}. The intrinsic anisotropy of some materials like phosphorene has shown intriguing phenomena such as negative reflection, anti-super-Klein tunneling, and perfect electronic waveguides \cite{BetancurOcampo2019, BetancurOcampo2020}. Meanwhile, the anisotropy induced by strain-engineering in graphene has led to the possibility of obtaining pseudo magnetic fields \cite{Levy2010, Guinea2009}, the tuning of electronic and optical properties \cite{Sena2012, Assili2015, DiazBautista2020, Ghosh2019, Pereira2009a, Pereira2009, Cocco2010, Pellegrino2010, Rostami2012, Barraza-Lopez2013, Naumis, Le2020, Cunha2020}, valley-polarized currents \cite{Stegmann2016, Stegmann2019}, and anomalous tunneling \cite{BetancurOcampo2018,BetancurOcampo2021}. 
The electronic and transport properties are valley-dependent due to the tilting of Dirac cones, as shown in the relative spacing of Landau levels \cite{Goerbig2008, Goerbig2009, Islam2017}, pseudo magnetic fields \cite{Zabolotskiy2016}, and other physical properties \cite{Morinari2009, Assili2015, Proskurin2015, Sari2015, Verma2017, Yang2018, Islam2018, Das2019, Jafari2019, Ohki2020, Menon2020, Sabsovich2020, Champo2019, Napitu2020,Faraei2020,Zheng2020,Sengupta2018,Zhang2018, Zhang2019,Zhou2020,Mojarro2021}. Recently, 8-$pmmn$ borophene and related tilted Dirac cone systems have been proposed as ideal solid-state platforms to realize analogs of gravitational waves in black holes and perform space-time engineering \cite{Farajollahpour2019, Farajollahpour2020}. These singular features motivate the investigation of unusual effects by the intrinsic Dirac cone tilt. Under the presence of uniform and crossed electric and magnetic fields, these outstanding electronic and transport properties may be tailored further.  

On the other hand, the Wigner function (WF) approach has been a powerful physical tool in quantum optics \cite{Lee1995, Case2008, Gerry2010, Baune2017, Rundle2017, Weinbub2018, 1977a}, which has also been applied recently in condensed matter in the study of electron dynamics in two-dimensional materials, particularly in graphene and under the presence of electromagnetic fields \cite{Weinbub2018, Jacoboni2004, Baeuerle2018, Colomes2015, Mason2015, Carmesin2020, Morandi2011, DiazBautista2020, DiazBautista2019, DiazBautista2017, Mason2013, Iafrate2017, Ghosh2021, Fernandez2020, Ferry2017}. The recent advances in the experimental reconstruction of the WF of electronic systems could bring the possibility of the realization of coherent electron states in the laboratory and the development of electron quantum optics \cite{Ferraro2014, Jullien2014, Baeuerle2018}. Thus, the research of new singular aspects in the recent discovery of two-dimensional materials, which contain anisotropic tilted Dirac cones, may offer new tracks for the design of protocols in a possible experimental setup of coherent states. For instance, in the context of valleytronics~\cite{Schaibley2016,Kundu2016}, the study of the density of states and conductivity of 8-$pmmn$ borophene under electric and magnetic fields evidenced a valley-dependence in magnetotransport properties and polarization currents~\cite{Islam2017,Islam2018}.  Weiss and Shubnikov-de Haas~\cite{LIFSHITS1958} oscillations have macroscopic manifestations which exhibit quantum and classical signatures simultaneously. In this manner, our general aim is to provide an adequate description in phase space of the physics of certain quantum phenomena, and their semi-classical representation, that occurs in condensed matter systems.

Thus, in this paper, we analyze how the Dirac cone tilt affects the time evolution of the WF of coherent electron states in the presence of crossed electric and magnetic fields. We find the exact solution of Landau states and energy spectra through a canonical transformation of the Dirac-Weyl Hamiltonian with tilted cones. The phase-space representation of Landau states evidences a valley dependency in the WF. The shape of the WF is strongly distorted and valley-dependent with the increasing of the electric field. We build coherent states in terms of the basis of the Hilbert space, which consists of the Landau states of the Dirac-Weyl Hamiltonian under electric and magnetic fields. The time evolution of these states and the WF show a significant temporal delay of electrons from one valley with respect to the other. In contrast to conventional coherent states, where the minimum uncertainty relation keeps constant, the uncertainties of position and momentum increase during the temporal evolution. Depending on the valley index, these coherent electron states can behave like coherent light states either for critical values of the electric field or the tilt parameter near the collapse of Landau levels. The mean trajectory of coherent states in the phase space is a spiral, and the WF presents more negative values going towards the origin. It agrees with the increase of position and momentum uncertainties in the time evolution. Such results may be not only tested by optical spectroscopy \cite{Sheng2019, Gao2020} or magnetotransport measurements \cite{Willke2017,ParaviciniBagliani2018}, but also by quantum tomography \cite{Jullien2014,Baeuerle2018,Takeda2021} for the reconstruction of the valley-dependent WF of tilted anisotropic Dirac materials.

\section{Electron dynamics in tilted anisotropic Dirac cone materials}\label{model}
\begin{figure}[t!!]

 \includegraphics[width=0.38\textwidth]{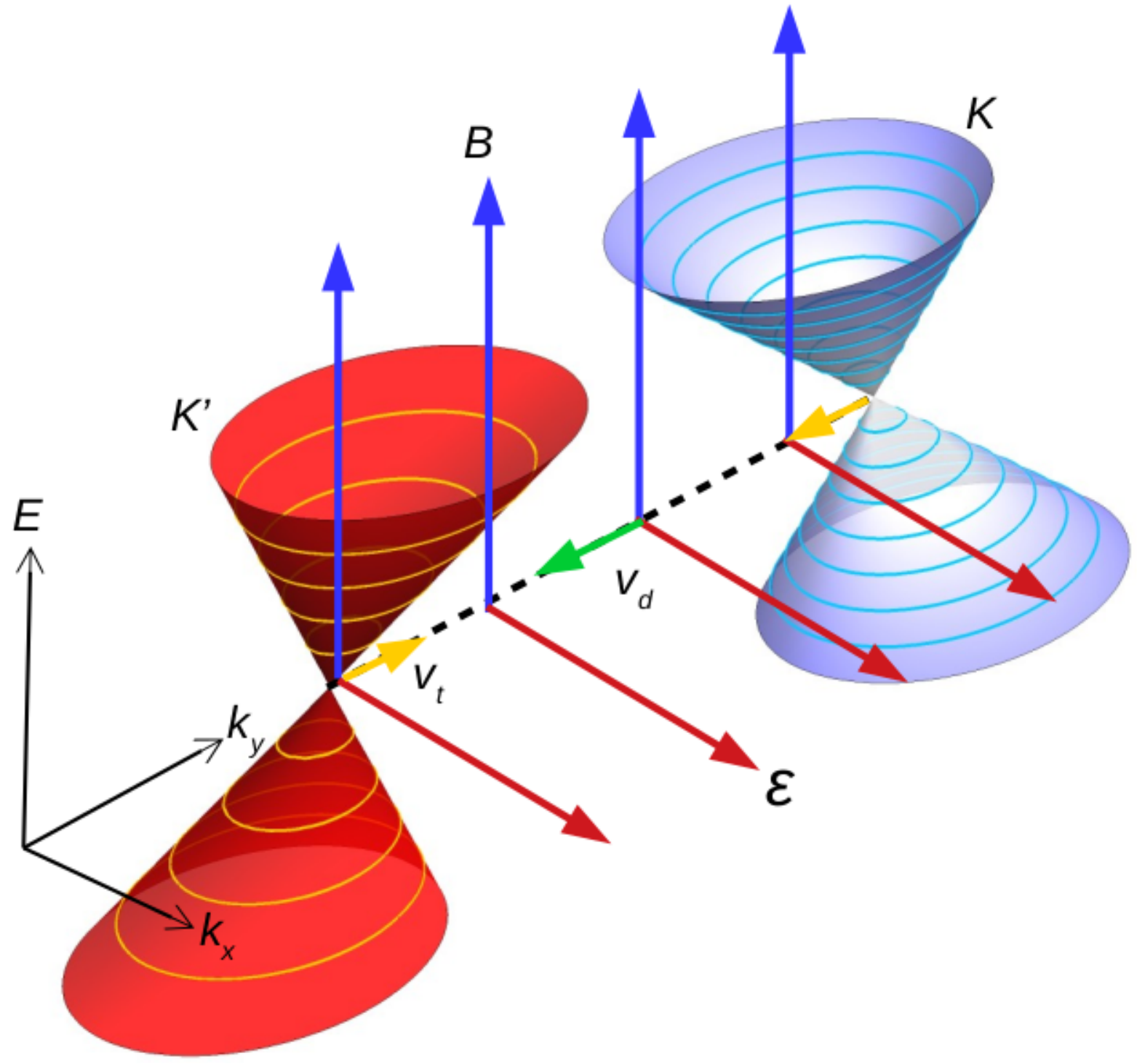} 
	\caption{\label{Cones} Electronic band structure of a 2D material with tilted anisotropic Dirac cones at the K and K' points under crossed electric $\vec{\mathcal{E}}$ and magnetic $\vec{B}$ fields. The green and yellow arrows correspond to drift $\vec{v}_{\rm d}$ and tilt $\vec{v}_t$ velocities, respectively. The energy isolines represent Landau levels, which are distributed differently in the two valleys, see Eq. \eqref{energy}.}
\end{figure}
Several two-dimensional materials such as 8-$pmmn$ borophene \cite{LopezBezanilla2016, Zabolotskiy2016, Islam2017, Islam2018}, strained graphene \cite{Assili2015}, Weyl semimetals \cite{Sabsovich2020, Das2019, Menon2020, Ferreira2021}, and the organic compound $\alpha$-(BEDT-TTF)$_2$I$_3$ \cite{Sari2015, Goerbig2008, Goerbig2009, Morinari2009} present a tilted Dirac cone in the low electronic band structure. The electronic properties of the tilted anisotropic Dirac materials are described by the continuous Dirac Hamiltonian

\begin{equation}\label{TAH}
    H = \nu\,(v_{t}\sigma_0 p_{y} + v_{x}\sigma_x p_{x} + v_{y}\sigma_y p_{y}),
\end{equation}
where $v_x$ and $v_y$ are the anisotropic Fermi velocities and $v_t$ is the tilt velocity quantifying the tilting of the Dirac cone. The time reversal symmetry operation $\vec{p}\rightarrow -\vec{p}$ allows interchanging the Dirac cones at valleys K and K'. It is possible to get different representations of the Hamiltonian in Eq. \eqref{TAH} depending on the election of the frame system and spinor \cite{Jafari2019, Goerbig2009}. These velocities depend on the material and geometrical restrictions exist for the ratio $v_t/v_y$ to change the dispersion relation. Since $v_t$ modulates the tilting of Dirac cones, when $v_t \geq v_y$ the Dirac cone intersects the plane $k_x$-$k_y$. For illustrative proposes and without loss of generality, in the forthcoming sections we will use the set of values $v_x = 0.86$, $v_y = 0.69$, and $v_t = 0.32$ in multiples of the Fermi velocity $v_F = 1$ that correspond to 8-$pmmn$ borophene~\cite{Zabolotskiy2016}. The matrices $\sigma_j$ with $j = x,\,y$ are the Pauli matrices, while $\sigma_0$ is the identity matrix. The quantity $\nu$ allows us to transit from valley K ($\nu = 1$) to valley K' $(\nu = -1)$. The eigenenergies of Hamiltonian \eqref{TAH} depict the low-energy excitations near the Fermi level as 
\begin{equation}
    E = \nu\,v_{t} p_{y} + \lambda\sqrt{v_{x}^2 p^2_x + v_{y}^2 p^2_y},
\end{equation}
which show an elliptical and tilted Dirac cone around valley K (see Fig.~\ref{Cones}). The band index $\lambda$ indicates the conduction ($\lambda = 1$) or ($\lambda = -1$) valence band. The electron dynamics at the other valley K' is described by Hamiltonian in Eq. \eqref{TAH} with a tilt velocity of opposite sign. Therefore, the Dirac cone at valley K' has opposite tilt compared to valley K (see Fig. \ref{Cones}).

To analyze the dynamics of electrons in tilted anisotropic Dirac materials under the presence of an in-plane electric field $\vec{\mathcal{E}} = \mathcal{E} \hat{x}$ and a perpendicular magnetic field $\vec{B} = B \hat{z}$, we include these fields in the Hamiltonian in Eq. \eqref{TAH} through their corresponding scalar and vector potentials, to get
\begin{equation}\label{EMH}
    H' = \nu\,[v_{x}\sigma_x p_{x} + (v_t\sigma_0 + v_{y}\sigma_y)(p_y + xB)] + x\mathcal{E}\sigma_0,
\end{equation}
where we use the minimal substitution with the Landau gauge $A_y = x B$. We set the direction of the electric field along with the $x$ axis to obtain the conservation of the linear momentum in $y$. For an arbitrary orientation, it is more convenient to choose the Landau gauge perpendicular to the electric field. In this case, the conserved linear momentum is $-p_x\sin\theta + p_y\cos\theta$, where $\theta$ is the angle of $\vec{\mathcal{E}}$ with the $x$ axis. This substitution is fulfilled if the magnetic length $l_{\rm B} = 1/\sqrt{B}\simeq26/\sqrt{B\,[\textrm{T}]}$ nm~\cite{Goerbig2009} is larger than the lattice constant of the crystal. In our calculation, we set the electron charge $e = -1$ and the Planck's constant $\hbar = 1$. The Hamiltonian in Eq. \eqref{EMH} has translational symmetry in the $y$ direction, and therefore, the linear momentum $p_{y} = k_y$ is conserved, while the application of the magnetic field $\vec{B}$ breaks the time-reversal symmetry. The conservation of $p_y$ allows using the following ansatz for the wave function:
\begin{equation}
\bar{\Psi}(x,y) = \exp(ik_{y}y)\Psi(x).
\end{equation}

\begin{figure*}[t!!]
  \begin{tabular}{ccc}
  (a) \qquad \qquad \qquad \qquad \qquad \qquad \qquad \qquad & (b) \qquad \qquad \qquad \qquad \qquad \qquad \qquad \qquad \\
 \includegraphics[width=0.38\textwidth]{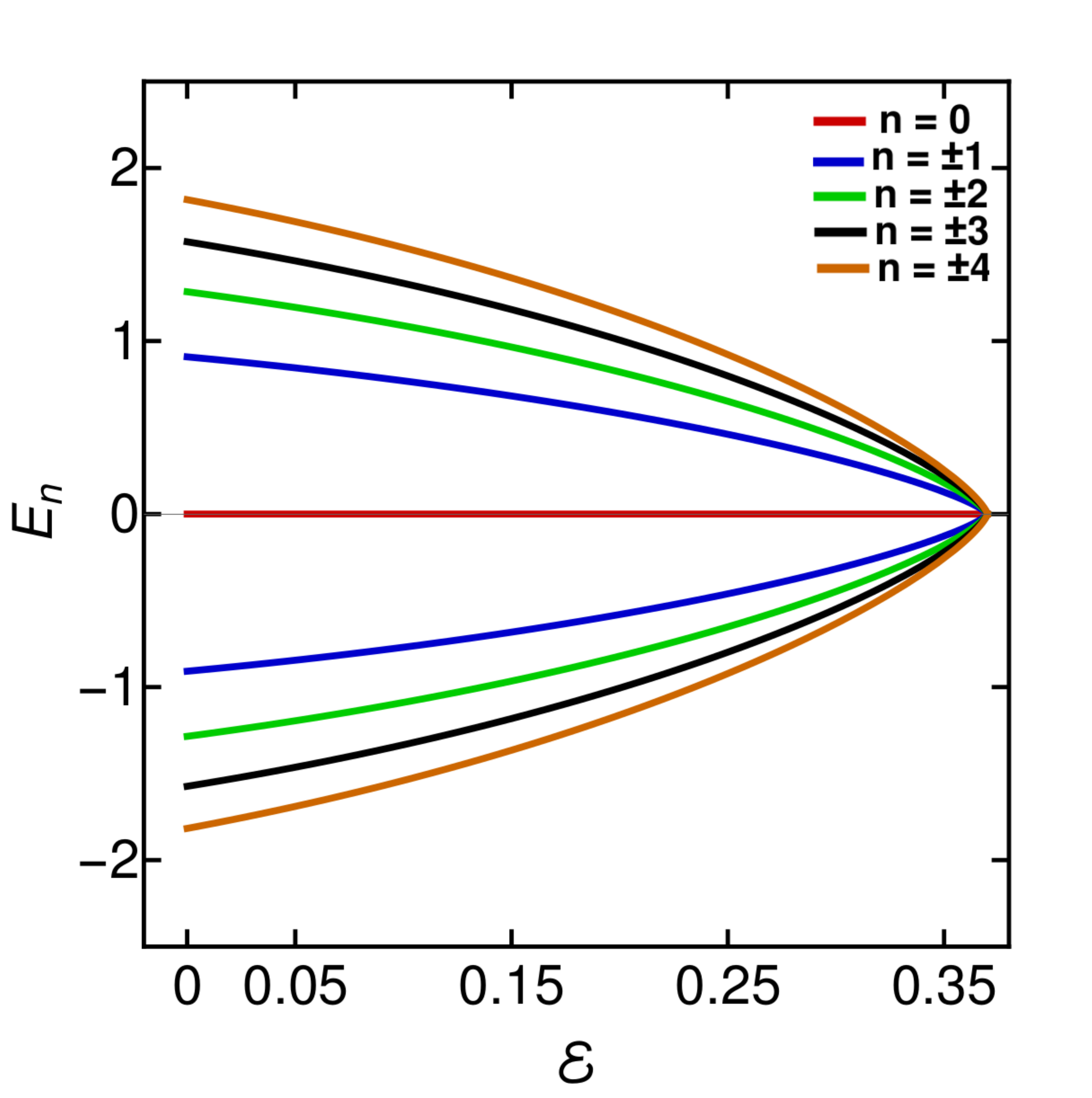} &
  \includegraphics[width=0.38\textwidth]{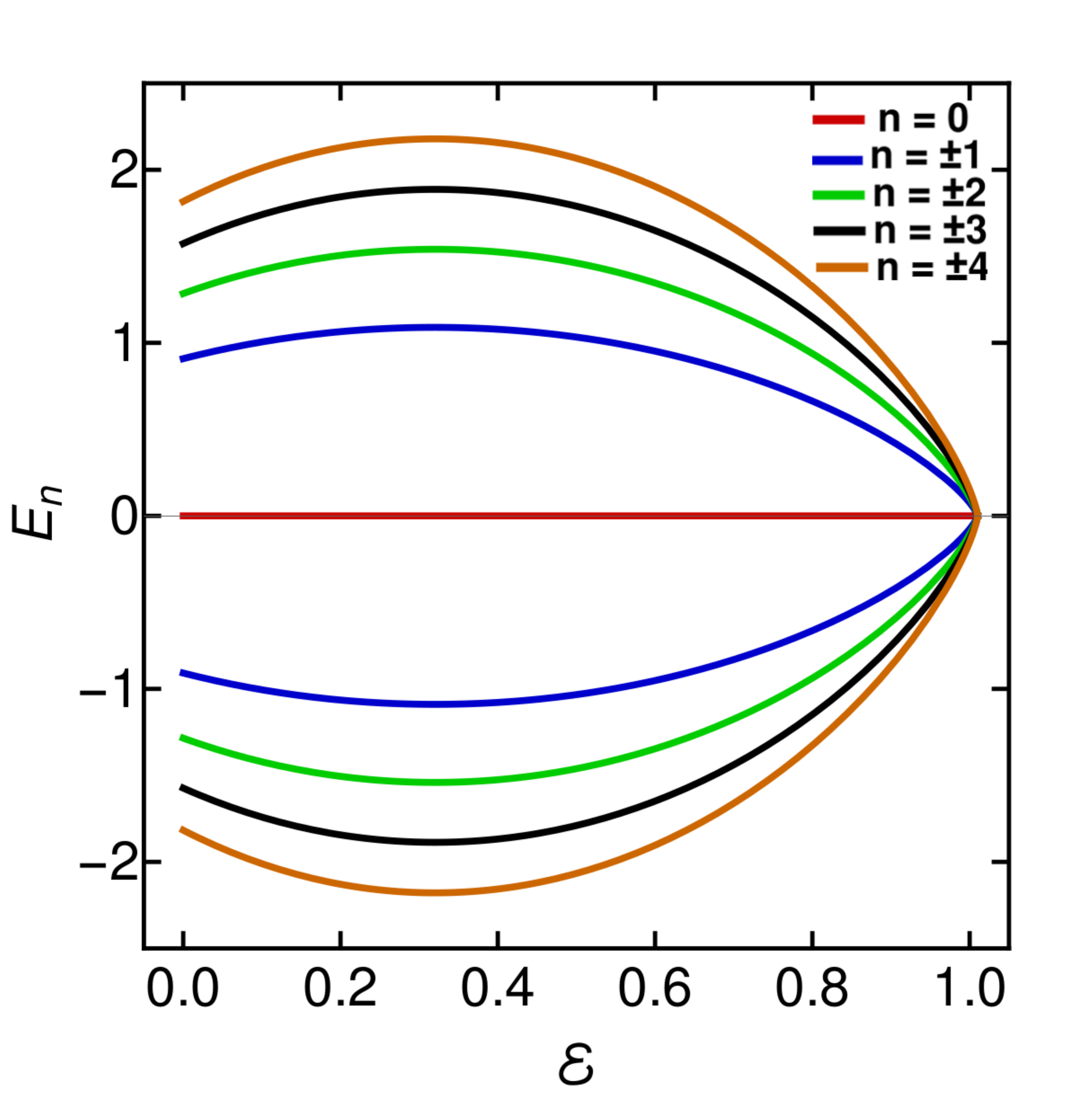} 
  \end{tabular}
	\caption{\label{spectrum} Energy spectrum in (\ref{energy}) with $k_{y}=0$ and $B = 1$ as a function of the electric field $\mathcal{E}$ for 8-$pmmn$ borophene at the K (a) and K' (b) points.}
\end{figure*}

\noindent We use the canonical transformation
\begin{subequations}
   \begin{eqnarray}
    x = x_c\sqrt{\frac{v_x}{v_y}}, & \qquad & y = y_c\sqrt{\frac{v_y}{v_x}},\\
    p_x = p^c_{x}\sqrt{\frac{v_y}{v_x}}, & \qquad & p_y = p^c_{y}\sqrt{\frac{v_x}{v_y}},
\end{eqnarray} 
\label{CT}
\end{subequations}
\noindent which lets invariant the commutation relations of position and linear momentum. With this transformation, the Hamiltonian in Eq. \eqref{EMH} is mapped to
\begin{equation}
    H'  =   \nu\,v'_{\rm F}\vec{\sigma}\cdot(\vec{p}_c + \vec{A}_c) + (\nu\,v_{t} p_{y} + x_c\overline{\mathcal{E}})\sigma_0 =  H'_0 + \nu\,v_{t} p_{y}\sigma_0,
    \label{Hc}
\end{equation}
\noindent where $v'_{\rm F} = \sqrt{v_x v_y}$ is the effective Fermi velocity, $\vec{A}_c = x_c B\hat{y}$ is the vector potential, and the Hamiltonian 
\begin{equation}\label{H'_0}
H'_0 = \nu\,v'_{\rm F}\vec{\sigma}\cdot(\vec{p}_c + \vec{A}_c) +  x_c\overline{\mathcal{E}}\sigma_0 
\end{equation}
\noindent describes the interaction of massless Dirac fermions with an effective electric field $\overline{\mathcal{E}} = (\mathcal{E} +\nu\, v_{t} B)\sqrt{v_x/v_y}$ and a magnetic field $B$ without the tilting of Dirac cones and anisotropy. Landau levels and bound states of Eq.~\eqref{H'_0} have been reported in Ref.~\cite{Lukose2007}, where a boost transformation makes zero the electrostatic potential in the Hamiltonian in Eq. \eqref{H'_0} and the magnetic field decreases with the factor $\sqrt{1 - (\mathcal{E}/v'_{\rm F} B)^2}$. In this way, the inverse boost transformation helps to solve the eigenenergies of Hamiltonian in Eq. \eqref{H'_0} exactly. This procedure allows us here to obtain the spectrum of the tilted anisotropic Dirac materials by using the inverse canonical transformation in Eqs.~\eqref{CT} (see Fig.~\ref{spectrum}):
\begin{equation}\label{energy}
E_{n,k_y}=\text{sgn}(n)\frac{\sqrt{v_x v_y}}{l_{\rm B}} (1-\beta^2_\nu)^{3/4}\sqrt{2|n|}-k_y\frac{\mathcal{E}}{B},
\end{equation}
where
\begin{equation}\label{beta}
\beta_\nu = \frac{\mathcal{E}}{v_{y} B} + \nu \frac{v_{t}}{v_{y}} = \frac{v_{\rm d} + \nu v_t}{v_y} 
\end{equation}

\noindent and $v_{\rm d} = \mathcal{E}/B$ is the drift velocity. Without loss of generality, we define $B = 1$ in all our calculations to express the electric field in units of the Fermi velocity $v_{\rm F}$. It is important to note that the Landau-level spectra in Eq. \eqref{energy} depends on valleys K and K' (see Fig.~\ref{spectrum}). This dependence of Landau levels with the valleys and the direction of the electric field was analyzed in Ref.~\cite{Goerbig2009}, where the maximum difference occurs when the electric field is perpendicular to the tilt velocity direction. In the case of an electric field oriented in an arbitrary direction, the valley dependence of the Landau-level spectra persists generally, because the parallel component of the drift velocity $\vec{v}_{\rm d}=\mathcal{\vec{\mathcal{E}}\times\vec{B}}/B^2$ to the tilt direction breaks the mirror symmetry of Dirac cones with respect to the $x$ axis. In the case when the electric field is parallel to the tilt direction, the drift velocity $\vec{v}_{\rm d}$ is perpendicular to the tilt velocity $\vec{v}_t$ and Landau levels are valley degenerated. The factor in Eq. \eqref{beta} indicates whether the orbits are closed when $|\beta_\nu| < 1$ or opened for $|\beta_\nu| \geq 1$ \cite{Lukose2007, Goerbig2009, Gu2011}. The transition from closed to opened orbits appears for the collapse of Landau levels at the critical value of $\beta_\nu = \pm 1$, where classically the electrons travel on a straight line, since  the drift and tilt velocities $v_{\rm d} + \nu v_{t}$ exceeds the velocity $v_y$~\cite{Goerbig2009}. Such a collapse can be reached tuning the electric field, the tilting of the Dirac cones, or the effective Fermi velocity $v'_{\rm F}$ to be zero \cite{Lukose2007, Morinari2009, Goerbig2008, Goerbig2009, Gu2011, Ghosh2019, Le2020}. For instance, in 8-$pmmn$ borophene, the collapse of Landau levels ($\beta_+ = 1$) occurs for valley K at the critical electric field or drift velocity $\mathcal{E}_c = v^c_{\rm d} = 0.37$, but this same value in valley K' corresponds to $\beta_{-} = 0.07$ and electrons have closed orbits. If we change the electric field to collapse the spectrum in valley K' ($\beta_{-} = 1$), this value is $\mathcal{E}_c = 1.01$, while in valley K the parameter $\beta_+ = 1.93$, indicating that the electrons escape of the magnetic confinement. The behavior of the Landau-level spectra in valleys is interchanged by reversing the critical electric field direction, which corresponds to the solution $\beta_\nu = -1$. It is important to note that the critical values of $v_{\rm d}^{c} = \mathcal{E}_{c}$ depend on each tilted anisotropic Dirac material, besides being constant amounts.

Now, to obtain the coherent states and their corresponding WF in the next sections, we write the eigenstates of the Hamiltonian in Eq. \eqref{EMH} as
\begin{equation}
 \Psi_n(x) =\mathbb{M}\Phi_{n}(x),
\label{LS}
\end{equation}
\noindent where 
	\begin{eqnarray}\label{M}
	\mathbb{M} & = & \sqrt{\frac{1}{2}}\left(\begin{array}{c c}
	\sqrt{C_+} & i\sqrt{C_-}\\
	-i\sqrt{C_-} & \sqrt{C_+}
	\end{array}\right), \nonumber\\ 
	\Phi_{n}(x)& = &\frac{1}{\sqrt{2^{(1-\delta_{0n})}}}\left(\begin{array}{c}
	(1-\delta_{0n})\psi_{n-1}(x) \\
	i\lambda\,\psi_{n}(x)
	\end{array}\right).
	\end{eqnarray}
The quantity $\delta_{mn}$ denotes the Kronecker delta, the entries of the matrix $\mathbb{M}$ are $C_{\pm}=1\pm\sqrt{1- \beta_\nu^2}$, and the components of the pseudo-spinor $\Phi_{n}(x)$ are given by the functions 
\begin{eqnarray}\label{5}
\psi_n(\xi_n) & = & \frac{(1- \beta_\nu^2)^{1/8}}{\sqrt{2^nn!\,l_{\rm B}}}\left(\frac{v_y}{\pi v_x}\right)^{1/4} \textrm{e}^{-\frac{1}{2}\xi^2_n} H_n\left(\xi_n\right), \nonumber\\
&&
\end{eqnarray}
where $H_n(\cdot)$ are the Hermite polynomials. The quantity $\xi_{n}$ is given by
\begin{equation}\label{xi}
\xi_{n}=(1- \beta_\nu^2)^{1/4}\sqrt{\frac{v_y}{v_x}}\left(\frac{x}{l_{\rm B}} + l_{\rm B} k_y\right) + \textrm{sgn}(n)\beta_\nu\sqrt{2|n|}.
\end{equation}
We note that for $\beta_\nu=0$ in both valleys, which corresponds to zero electric field and tilt, the eigenspinors $\Phi_{n}$ in \eqref{5} reduce to the solutions of the Landau-level spectra and states of anisotropic massless Dirac fermions \cite{DiazBautista2020}. 

\section{Coherent electron states}\label{annihilation}
Coherent states arise in multiple branches of physics, mainly in quantum optics and information processes \cite{Weinbub2018, Gerry2010}. The minimal uncertainty of these states makes them the most classical states in quantum mechanics. In condensed matter, coherent states can be observed in low-temperature phenomena, such as superconductivity~\cite{Bardeen1957,Anderson1958,Hofheinz2008,Siddiqi2021} and Bose-Einstein condensates~\cite{Greiner2002,Berrada2013,Hashimoto2020,Pirro2021}. The resemblance of electrons in 2D materials with photons, mainly due to the linear dispersion relation, leads us to consider the possibility of obtaining coherent states in electronic systems. The presence of crossed electric and magnetic fields as well as the tilting of Dirac cones of the material are important keys for the quantization of energy spectra. Moreover, the application of electric and magnetic fields offers additional degrees of freedom to manipulate the trajectory of electrons. For these purposes, we build the coherent states defining an annihilation operator $\Theta_{n}^{-}$ that acts onto the Hilbert basis $\Phi_n(\vec r)$. With the inverse matrix $\mathbb{M}^{-1}$, we obtain from~\eqref{LS} \cite{celeita2020}
\begin{equation}\label{31}
\Phi_{n}(x,y)=\mathbb{M}^{-1}\Psi_{n}(x,y).
\end{equation}

\noindent We also define the differential operators
	\begin{equation}
	\theta_{n}^{\pm}=\frac{1}{\sqrt{2}}\left(\mp\frac{d}{d\xi_{n}}+\xi_{n}\right), \quad \theta_{n}^{+}=(\theta_{n}^{-})^{\dagger},
	\end{equation}
	and the unitary shift operators $\mathcal{T}^{\pm}$~\cite{Golinski2019}, 
	whose action onto the eigenfunctions $\psi_{n}(\xi_{n})$ is
	\begin{eqnarray}
	\mathcal{Q}^{-}\psi_{n}(\xi_{n})& =& \mathcal{T}^{-}\theta_{n}^{-}\psi_{n}(\xi_{n})\equiv\sqrt{n}\,\psi_{n-1}(\xi_{n-1}), \nonumber \\
	\mathcal{Q}^{+}\psi_{n}(\xi_{n})& =& \theta_{n}^{+}\mathcal{T}^{+}\psi_{n}(\xi_{n})\equiv\sqrt{n+1}\,\psi_{n+1}(\xi_{n+1}). \nonumber \\
	&&
	\end{eqnarray}
Therefore, the operators $\theta_{n}^{\pm}$ lower and raise the eigenstates $\psi_{n}$, while $\mathcal{T}^{\pm}$ shifts the index $n$ of the spatial coordinate $\xi_{n}$ by a unity. Moreover, we can verify that $[\mathcal{Q}^{-},\mathcal{Q}^{+}]=1$.

In this way, we build the following operators:
\begin{subequations}\label{ladder}
\begin{align}
    &\Theta_{n}^{-}  = \frac{1}{\sqrt{2}}\left(\begin{array}{c c}
    \frac{\sqrt{N+2}}{\sqrt{N+1}}\mathcal{Q}^{-} & -i\,\lambda\mathcal{T}^{+}\frac{1}{\sqrt{N+1}}(\mathcal{Q}^-)^2 \\
    i\,\lambda\mathcal{T}^{-}\sqrt{N+1} & \mathcal{Q}^{-}\end{array}\right),\\
    &\Theta_{n}^+  =  (\Theta_{n}^-)^\dagger,
\end{align}
\end{subequations}
\noindent  in terms of $\mathcal{Q}^{\pm}$~\cite{DiazBautista2017, DiazBautista2019}, where $N=\mathcal{Q}^+\mathcal{Q}^-$ is the number operator. The index of the matrix operators in~\eqref{ladder} means that for each eigenvector $\Psi_{n}$, there is a set of ladder operators with the same index. To avoid this, we join such matrix operators with a one-dimensional projection operator $\mathcal{P}(k)$~\cite{Isozaki2011}. Thus, we are able to define two ladder operators as
	\begin{equation}
	\Theta^{-}\equiv\sum_{n=0}\Theta_{n}^{-}\mathcal{P}(n), \quad \Theta^+=(\Theta^-)^\dagger,
	\end{equation}
such that
\begin{equation}
	\Theta^{\pm}\Phi_{k}\equiv\sum_{n=0}\Theta_{n}^{\pm}(\mathcal{P}(n)\Phi_{k})=\sum_{n=0}\delta_{kn}\Theta_{n}^{\pm}\Phi_{k}=\Theta_{k}^{\pm}\Phi_{k}.
\end{equation}
	
Hence, the actions of the ladder operators $\Theta^{\pm}$ on the Hilbert states $\Phi_{n}(x,y)$ are given by
\begin{subequations}
\begin{align}\label{35}
	\Theta^-\Phi_{n}(\xi_{n},y)&=  \sqrt{2^{(1-\delta_{1n})}}\sqrt{n}\Phi_{n-1}(\xi_{n-1},y), \\
	\Theta^+\Phi_{n}(\xi_{n},y)&=  \sqrt{2^{(1-\delta_{0n})}}\sqrt{n+1}\Phi_{n+1}(\xi_{n+1},y),
\end{align}
\end{subequations}
whose commutation relation reads
\begin{equation}
    [\Theta^{-},\Theta^{+}]\Phi_{n}(x,y)=c(n)\Phi_{n}(x,y), \quad c(n)=\begin{cases}
    1, & n=0 \\
    3, & n=1 \\
    2, & n>1.
    \end{cases}
\end{equation}

Now, we define the coherent states as eigenstates of the annihilation operator $\Theta^{-}$:
\begin{equation}\label{QO}
\Theta^-\Phi_{z}(x,y)=z\Phi_{z}(x,y), \quad z\in\mathbb{C},
\end{equation}
with complex eigenvalue, where
\begin{equation}
\Phi_{z}(x,y)=\sum_{n=0}^{\infty}a_{n}\Phi_{n}(x,y).
\end{equation}
Using Eq. \eqref{35}, the explicit expression for the coherent states is given by
\begin{eqnarray}\label{40}
\Phi_{z}(x,y) = \frac{\left[\Phi_{0}(x,y)+\sum_{n=1}^{\infty}\frac{\sqrt{2}\alpha^n}{\sqrt{n!}}\Phi_{n}(x,y)\right]}{\sqrt{2\exp\left(\vert \alpha\vert^2\right)-1}},
\end{eqnarray}
where $\alpha=z/\sqrt{2}=\vert\alpha\vert\exp\left(i\varphi\right)$. It is worth noting that the phase angle of $\alpha$ is identical to the angular rotation in the classical motion and carries information about cyclic changes for the average observable in the position and momentum \cite{Carruthers1965, Noel1995}. 

To analyze the time-dependent electron dynamics within the WF approach, it is necessary to apply the time evolution operator $U(t) = \exp(-i H t)$ on the expansion of coherent states in terms of Landau states $\Psi_n(x,y)$,
\begin{eqnarray}\label{timecoherentstate}
\Psi_{\alpha}(x,y,t)& = &\frac{1}{\sqrt{2\exp\left(\vert \alpha\vert^2\right)-1}}\mathbb{M}\left(\begin{array}{c}
\psi_{\alpha,1}(x,y,t)\\
i\,\lambda\,\psi_{\alpha,2}(x,y,t)
\end{array}\right), \nonumber\\
&&
\end{eqnarray}
where
\begin{subequations}
    \begin{eqnarray}
    \psi_{\alpha,1}(x,y,t)&=&\sum_{n=1}^{\infty}\frac{\alpha^{n}\textrm{e}^{-iE_n t}}{\sqrt{n!}}\psi_{n-1}(x,y), \\
    \psi_{\alpha,2}(x,y,t)&=&\sum_{n=0}^{\infty}\frac{\alpha^{n}\textrm{e}^{-iE_n t}}{\sqrt{n!}}\psi_{n}(x,y).
\end{eqnarray}
\end{subequations}

\begin{figure*}[p!!]
\begin{tabular}{c}
\includegraphics[trim = 0mm 0mm 0mm 0mm, scale= 0.21, clip]{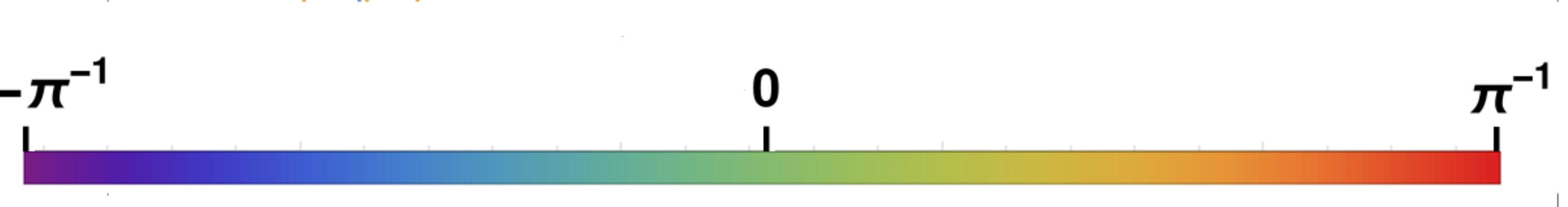}
\end{tabular}
\begin{tabular}{ccc}
(a) \qquad \qquad \qquad $n=0$, $\nu=1$, and $\mathcal{E} = 0.25 $\qquad \qquad \qquad & \qquad \qquad \qquad &(b)\qquad \qquad \qquad $n=0$, $\nu=-1$, and $\mathcal{E} = 0.25 $\qquad \qquad \qquad\\
\includegraphics[trim = 0mm 0mm 0mm 0mm, scale= 0.20, clip]{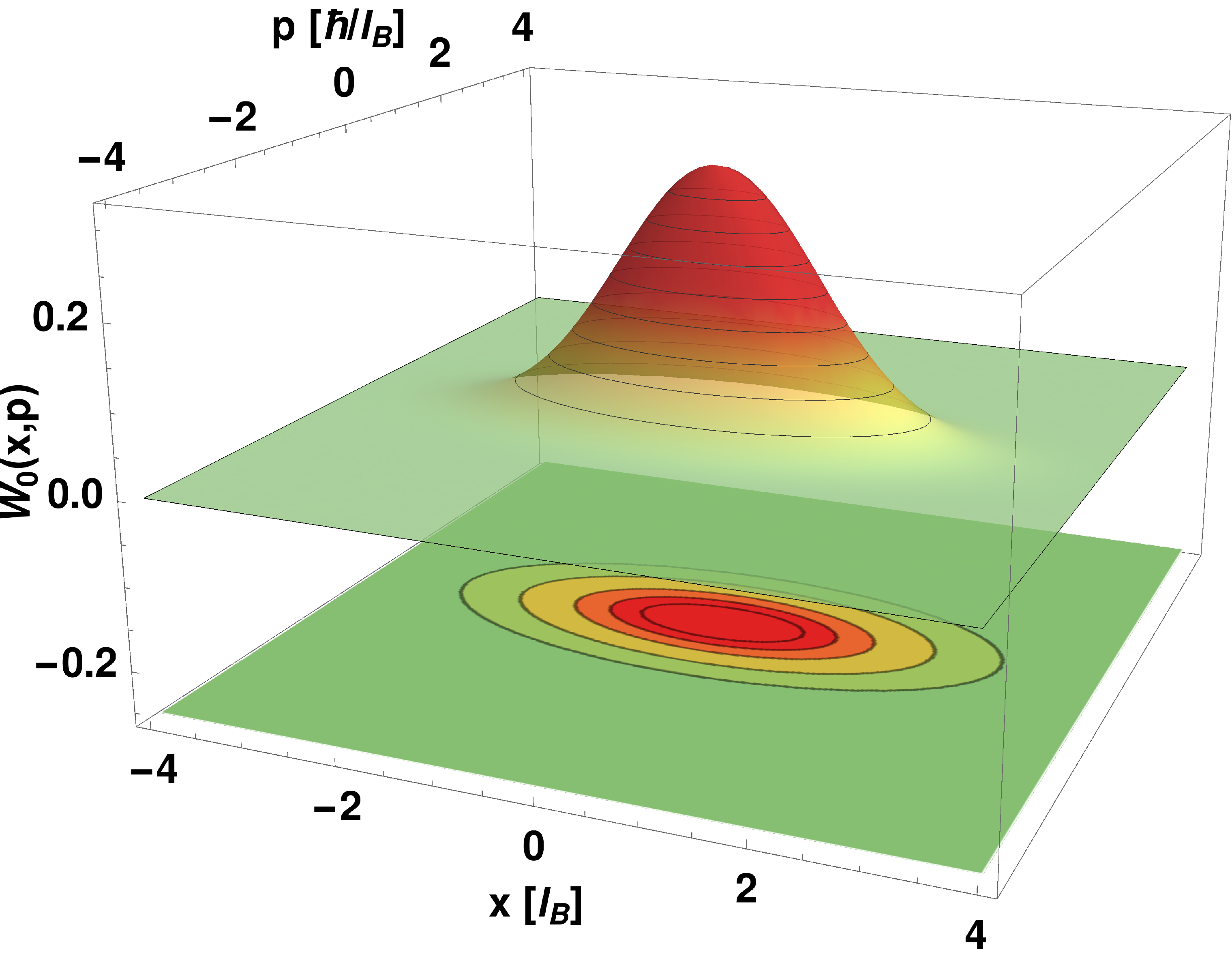} & \qquad \qquad \qquad &
\includegraphics[trim = 0mm 0mm 0mm 0mm, scale= 0.20, clip]{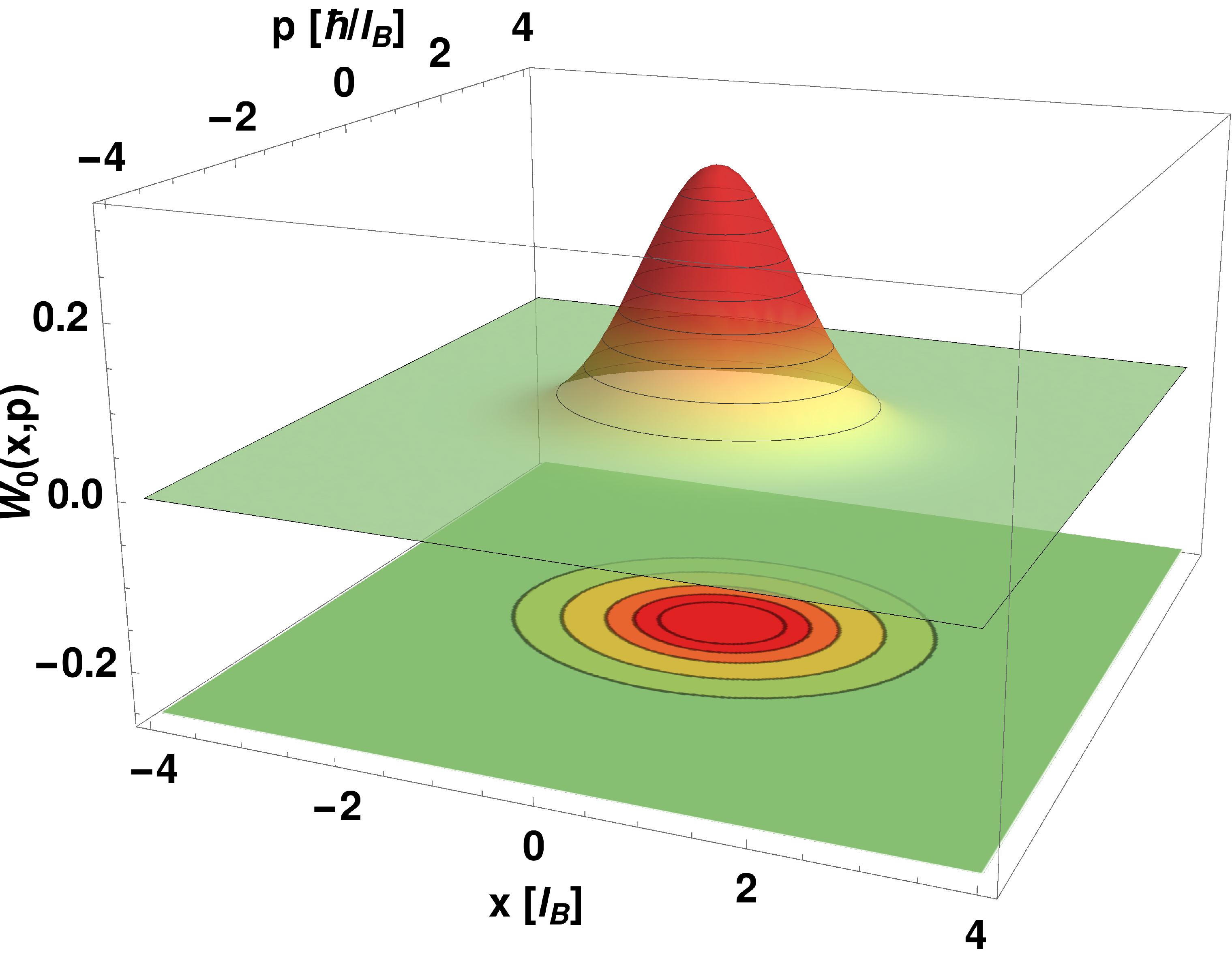}\\
(c) \qquad \qquad \qquad $n=1$, $\nu=1$, and $\mathcal{E} = 0.25 $\qquad \qquad \qquad & \qquad \qquad \qquad & (d) \qquad \qquad \qquad $n=1$, $\nu=-1$, and $\mathcal{E} = 0.25 $\qquad \qquad \qquad\\
\includegraphics[trim = 0mm 0mm 0mm 0mm, scale= 0.20, clip]{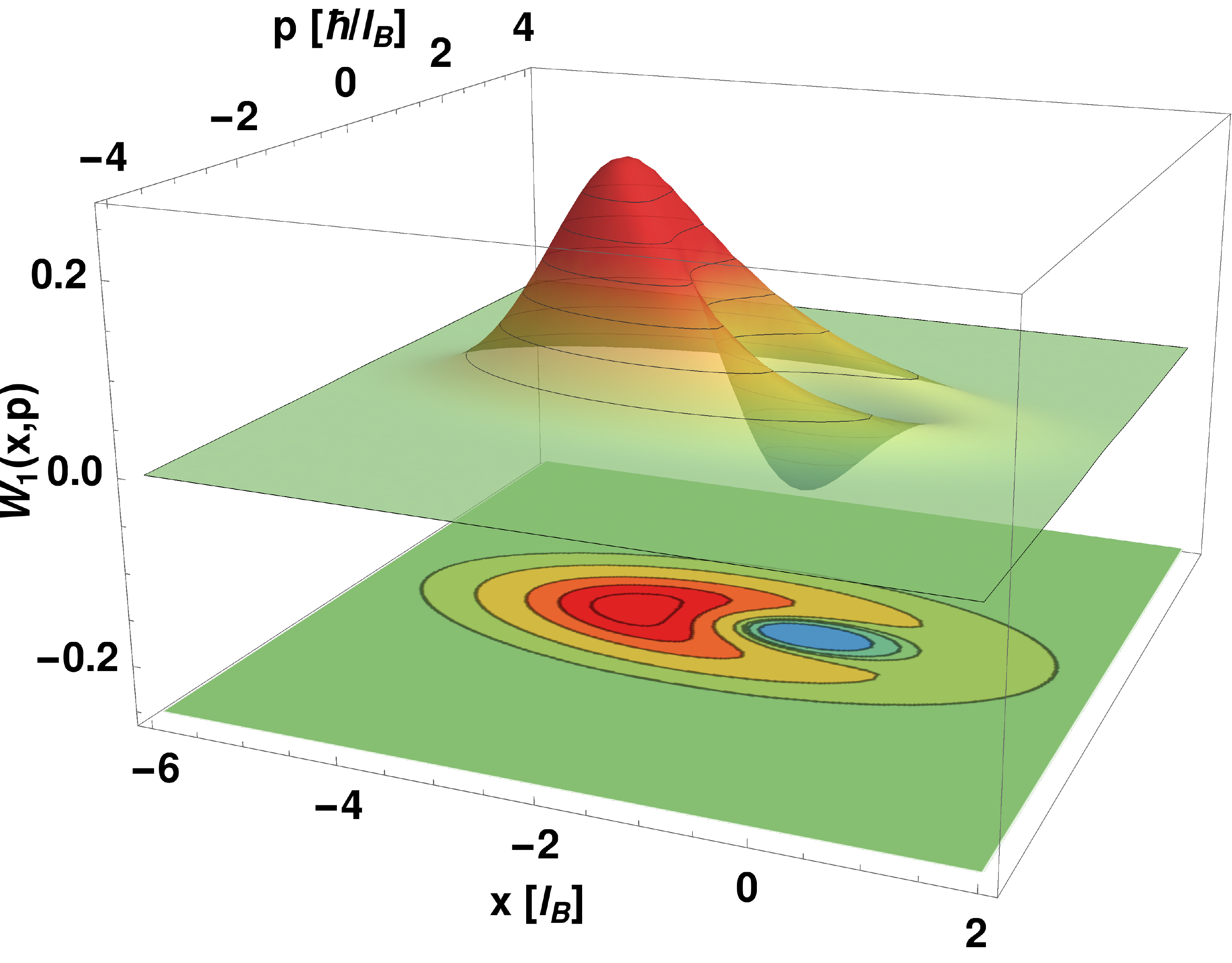} & \qquad \qquad \qquad &
\includegraphics[trim = 0mm 0mm 0mm 0mm, scale= 0.20, clip]{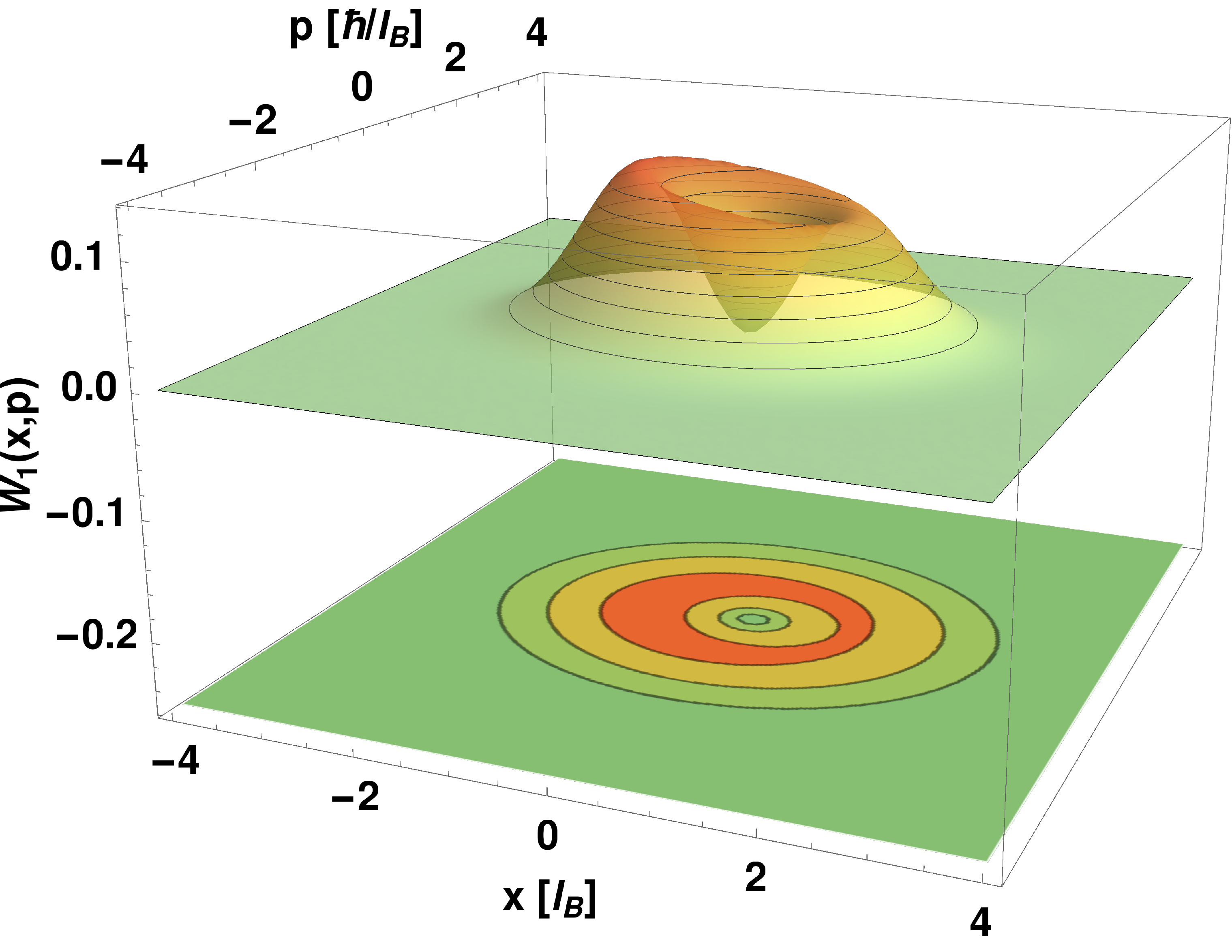}\\
(e) \qquad \qquad \qquad $n=2$, $\nu=1$, and $\mathcal{E} = 0.25 $\qquad \qquad \qquad & \qquad \qquad \qquad & (f) \qquad \qquad \qquad $n=2$, $\nu=-1$, and $\mathcal{E} = 0.25 $\qquad \qquad \qquad\\
\includegraphics[trim = 0mm 0mm 0mm 0mm, scale= 0.20, clip]{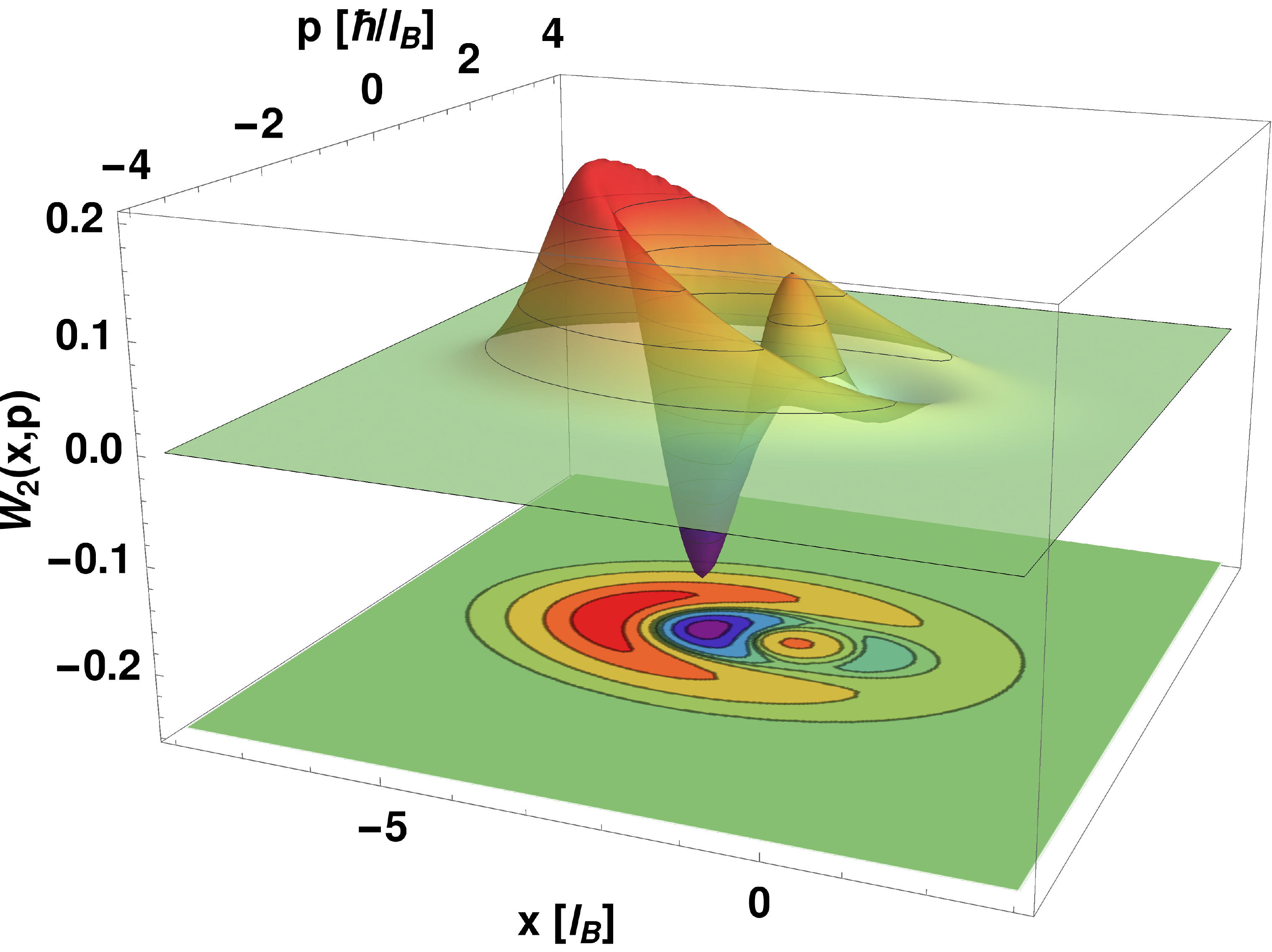} & \qquad \qquad \qquad &
\includegraphics[trim = 0mm 0mm 0mm 0mm, scale= 0.20, clip]{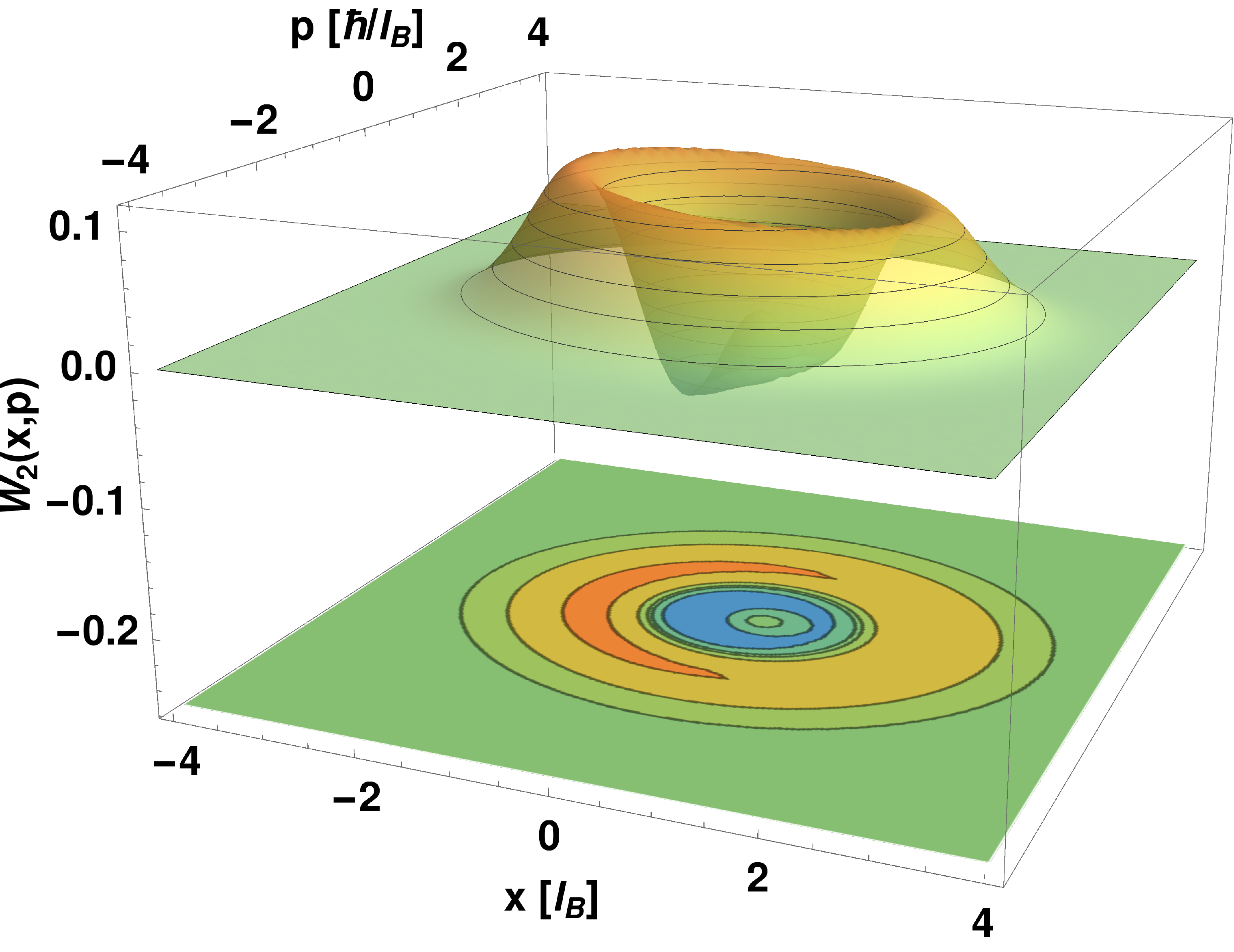}\\
(g) \qquad \qquad \qquad $n=2$, $\nu=1$, and $\mathcal{E} = 0 $ \qquad \qquad \qquad & \qquad \qquad \qquad & (h) \qquad \qquad \qquad $n=2$, $\nu=-1$, and $\mathcal{E} = 0. $ \qquad \qquad \qquad\\
\includegraphics[trim = 0mm 0mm 0mm 0mm, scale= 0.20, clip]{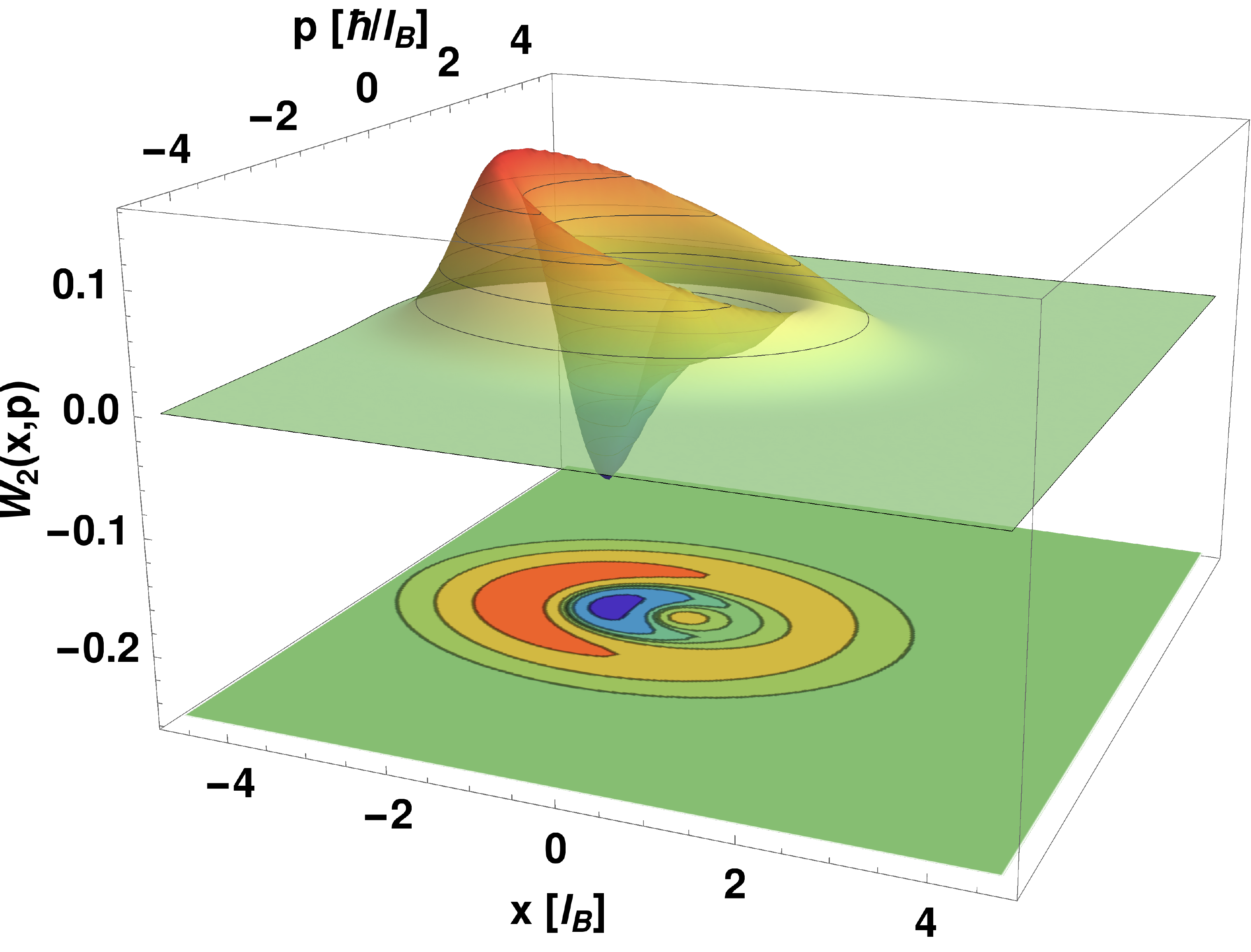} & \qquad \qquad \qquad &
\includegraphics[trim = 0mm 0mm 0mm 0mm, scale= 0.20, clip]{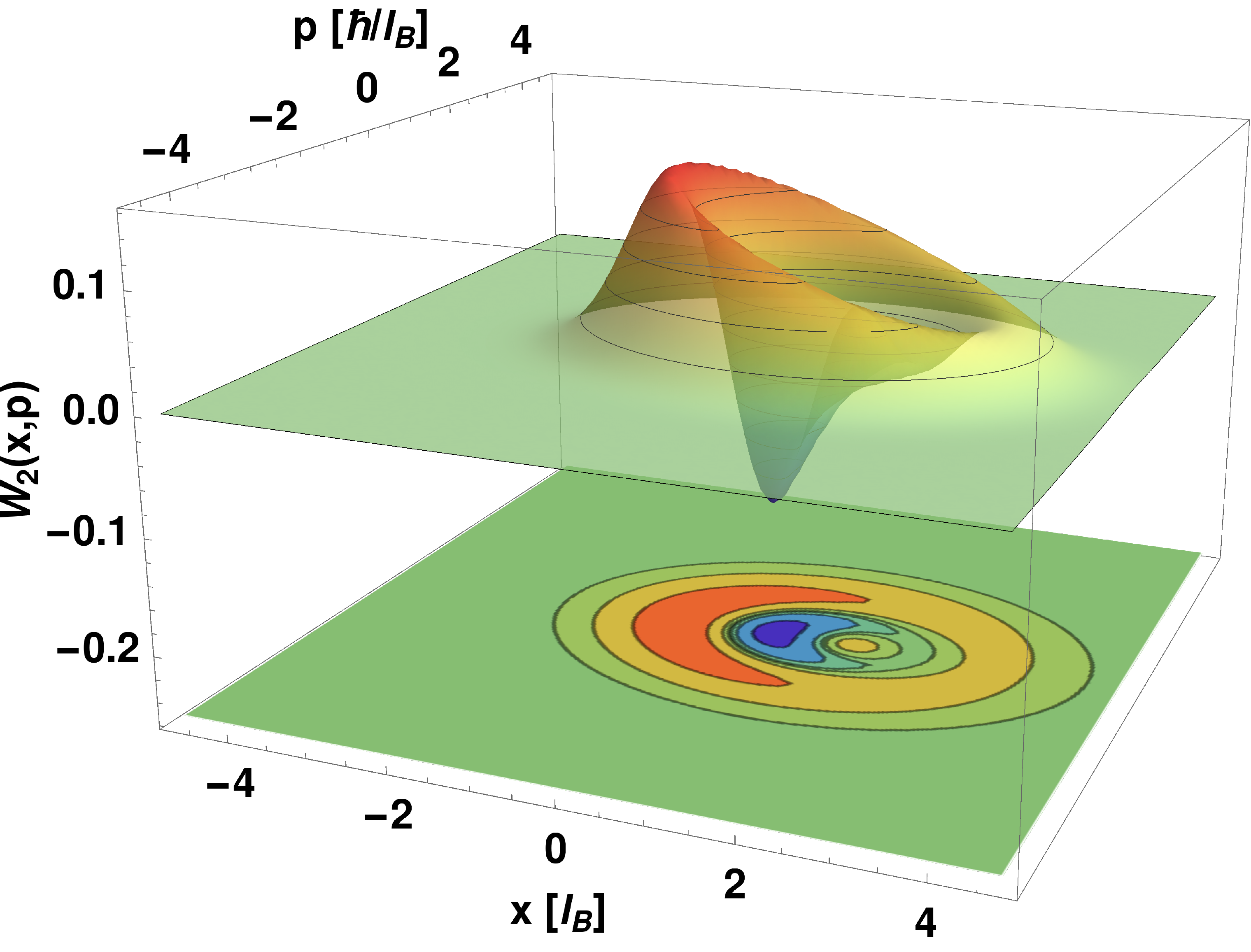}
\end{tabular}
\caption{Trace of the Wigner matrix $\mathbb{W}_{n}(\vec{r},\vec{p})$ in Eq.~(\ref{traceWn}) for different values of $n$ in each Dirac point ($\nu=\pm1$). We set the values of $\mathcal{E} = 0.25$ from (a)-(f), and $\mathcal{E} = 0$ for (g) and (h), $B=1$, $k_{y}=0$, $v_x = 0.86$, $v_y = 0.69$, $v_t = 0.32$, and $\lambda=1$.}
\label{fig2}
\end{figure*}

\section{Wigner function for Landau and coherent electron states}
The WF $W(\vec{r},\vec{p})$ is a quasiprobability distribution defined as \cite{Wigner1932, Hillery1984, Kenfack2004}
\begin{equation}
W(\vec{r},\vec{p})=\frac{1}{\left(2\pi\right)^n}\int_{-\infty}^{\infty}\textrm{e}^{i\,\vec{p}\cdot\vec{r}'}\left\langle\vec{r}-\frac{\vec{r}'}{2}\right\vert\rho\left\vert\vec{r}+\frac{\vec{r}'}{2}\right\rangle d\vec{r}',
\label{WM}
\end{equation}
where $\rho$ is the density matrix; $\vec{r}=(r_1,r_2,\dots,r_n)$ and $\vec{p}=(p_1,p_2,\dots,p_n)$ are $n$-dimensional vectors representing the classical phase-space position and momentum values, respectively; and $\vec{r}'=(r'_1,r'_2,\dots,r'_n)$ is the position vector in the integration process. In contrast with the probability density of any quantum state, the WF can take negative values. This negativity in the WF indicates the nonclassicality of a state, and it is interpreted as a sign of quantumness \cite{Smithey1993, Baune2017, Weinbub2018}. However, the probability distributions $|\psi(x)|^2$ and $|\phi(p)|^2$ can be obtained, integrating the WF on $x$ or $p$, as well as the normalization condition~\cite{Case2008}.

To calculate the Wigner matrix (WM) for the $n$-th Landau state in valleys K and K', we substitute the eigenstates in Eq. \eqref{LS} into the Wigner representation in Eq. \eqref{WM}:
\begin{equation}
\mathbb{W}_{n}(\vec{r},\vec{p})=\mathbb{M}W_{n}(\vec{r},\vec{p})\mathbb{M}^{\dagger},
\end{equation}
with
\begin{widetext}
\begin{equation}\label{6}
W_{n}(\vec{r},\vec{p})=\frac{W(y,p_{y})}{2^{(1-\delta_{0n})}}\left(\begin{array}{c c}
(1-\delta_{0n})W_{n-1,n-1}(x,p_{x}) & -i\,\lambda(1-\delta_{0n})W_{n-1,n}(x,p_{x}) \\
i\,\lambda(1-\delta_{0n})W_{n,n-1}(x,p_{x}) & W_{n,n}(x,p_{x})
\end{array}\right),
\end{equation}
\end{widetext}
where the components $W_{j,g}(x,p_{x})$ and $W(y,p_{y})$ are given by
\begin{subequations}
	\begin{align}
	W_{j,g}(x,p_{x})&=\frac{1}{\pi}\int_{-\infty}^{\infty}\textrm{e}^{2i\,p_{x}q_1}\psi_{j}(x-q_1)\psi_{g}^\ast(x+q_1) dq_1, \label{18} \\
	W(y,p_{y})&=\frac{1}{\pi}\int_{-\infty}^{\infty}\textrm{e}^{2i\left(p_{y}-k_{y}\right)q_2}dq_2\equiv\delta\left(p_{y}-k_{y}\right),
	\end{align}
\end{subequations}
\noindent being $j$ and $g$ positive integers. The wave functions $\psi_{j}$ and $\psi_{g}$ are the components given by \eqref{5}.

For computing the function $W_{j,g}(x,p_{x})$, we define the following quantities:
\begin{equation}\label{19}
u = (1- \beta_\nu^2)^{1/4}\frac{q_1}{l_{\rm B}}\sqrt{\frac{v_y}{v_x}}, \quad s=(1- \beta_\nu^2)^{-1/4}p_xl_{\rm B}\sqrt{\frac{v_x}{v_y}}.
\end{equation}
Hence, using these definitions and by substituting Eq.~(\ref{5}) in Eq.~(\ref{18}), we get
\begin{eqnarray}\label{38}
W_{j,g}(\chi_{n,m}) = \frac{1}{\pi}\exp\left(-\frac{1}{2}|\chi_{n,m}|^2+i(\xi_{n}-\xi_{m})s\right)&&\nonumber\\
\times\left\{\begin{array}{c c}
(-1)^j\sqrt{\frac{j!}{g!}}\chi_{n,m}^{g-j}L_{j}^{g-j}\left(|\chi_{n,m}|^2\right), & \text{if } j\leq g \\
(-1)^g\sqrt{\frac{g!}{j!}}\chi_{n,m}^{*j-g}L_{g}^{j-g}\left(|\chi_{n,m}|^2\right), & \text{if } j\geq g,
\end{array}\right. &&
\end{eqnarray}
where the functions $L_{n}^{m}(\cdot)$ are the associated Laguerre polynomials and the definition
\begin{equation}
\chi_{n,m} = \sqrt{2}\left(\frac{\xi_{n}+\xi_{m}}{2} +is\right).
\end{equation}

\noindent Thus, the components of the $2\times2$ WM turn out to be
\begin{subequations}\label{23}
	\begin{align}
	W_{n-1,n-1}(\chi_{n})&=\frac{1}{\pi}(-1)^{n-1}\textrm{e}^{-\frac{1}{2}|\chi_{n}|^2}L_{n-1}\left(|\chi_{n}|^2\right), \\
\nonumber	W_{n-1,n}(\chi_{n})&=\frac{(-1)^{n-1}}{\pi\sqrt{n}}\chi_{n}\textrm{e}^{-\frac{1}{2}|\chi_{n}|^2}L_{n-1}^{1}\left(|\chi_{n}|^2\right) \\
	&=W_{n,n-1}^\ast(\chi_{n}),\\
	W_{n,n}(\chi_{n})&=\frac{1}{\pi}(-1)^{n}\textrm{e}^{-\frac{1}{2}|\chi_{n}|^2}L_{n}\left(|\chi_{n}|^2\right),
	\end{align}
\end{subequations}
\noindent where $\chi_{n}\equiv\chi_{n,n}$. In this way, the trace of the WF $\mathbb{W}(\vec{r},\vec{p})$ is given by (see Fig.~\ref{fig2})
\begin{eqnarray}\label{traceWn}
{\rm Tr}[\mathbb{W}_{n}(\vec{r},\vec{p})]&=&\frac{\delta\left(p_{y}-k_{y}\right)}{2^{1-\delta_{0n}}}\Big\{W_{n,n}(\chi_{n})+ (1-\delta_{0n})\nonumber\\
& \times & \left[W_{n-1,n-1}(\chi_{n}) -2\lambda\beta_\nu\Re\left(W_{n-1,n}(\chi_{n})\right)\right]\Big\}, \nonumber\\
&&
\end{eqnarray}
where $\Re(z)$ denotes the real part of a complex number $z$. To illustrate the valley dependency of the WF for Landau level states in Eq. \eqref{traceWn}, we show the Landau states $n = 0, 1,$ and $2$ for valleys K and K' in Fig.~\ref{fig2}. Singular features of the WF of Landau states emerge by the combination of tilted Dirac cones and an electric field. The term $\textrm{sgn}(n)\beta_{\nu}\sqrt{2|n|}$ in Eq. \eqref{xi} determines the center of the WFs when $k_y = 0$, whose origin is due to the tilting of Dirac cones and the electric field. For the Landau level $n = 0$, the WF in both valleys appears at $x = 0$. But with $n$ different to zero, the WF for K and K' are located asymmetrically in the $x$ axis, as shown in Figs. \ref{fig2}(a)-(f). Importantly, this shift of the WF for Landau states still occurs in the absence of the electric field, see Figs. \ref{fig2}(g) and (h). WFs for $\nu = \pm 1$ have the same shape, but the center appears at symmetrical points given by $\textrm{sgn}(n)\nu v_t\sqrt{2|n|}/v_y$ (see Eqs.~(\ref{beta}) and (\ref{xi})). These features of the WF may be revealed through a quantum tomography experiment, even in the absence of an electric field, which is not possible by magnetoresistance and density of states measurements due to the valley degeneracy of Landau levels.

The effect of the electric field causes asymmetry in the WFs of both valleys due to the competition between drift and tilt velocities, as shown in Figs. \ref{fig2}(a)-(f). Since $\beta_+ > \beta_-$ for valleys K and K', the WF of the Landau-level spectra with $\nu = 1$ is more distorted than the one with $\nu = -1$. For the degenerated level $n = 0$, we can observe that the shape of the WF becomes different for each valley. The electric field scales the position and linear momentum differently by the opposite tilting of the Dirac cones in the different valleys. From the critical electric field $\mathcal{E}_c = (v_{y}-\nu v_{t}) B$, Landau levels and WF collapse, since physically the system is in a regime in which the electric field dominates over the magnetic field. Therefore, it is possible to get open orbits of electrons, while in the other valley, electrons present closed trajectories with discrete energy spectra represented by Landau levels in Eq. \eqref{energy}.

\begin{figure}[ht]
	\centering
	\includegraphics[width=8cm]{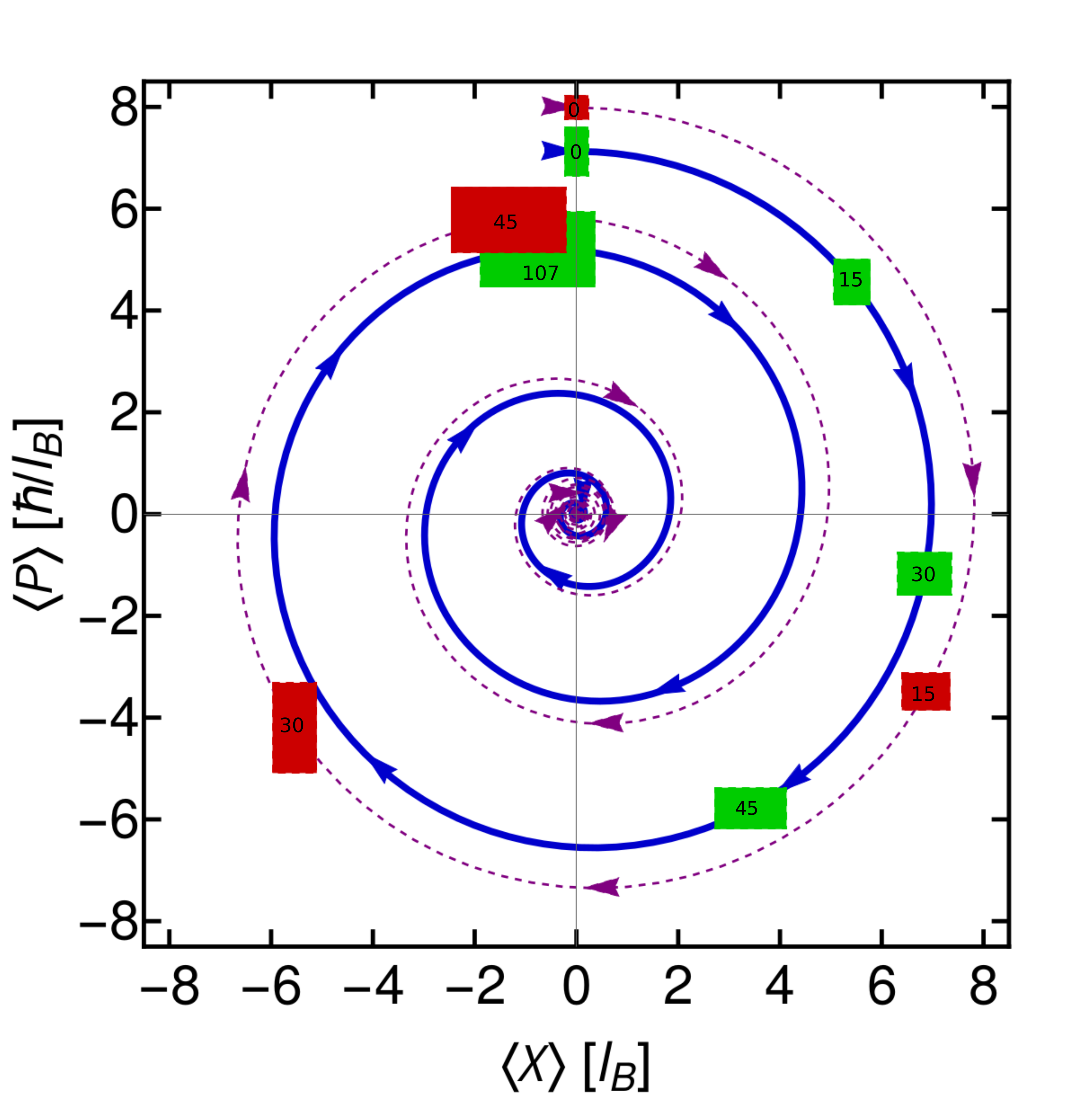}
	\caption{\label{fig:spiral} Mean trajectory of the coherent states from the expectation values in Eq. \eqref{EQO} of the operators $X$ and $P$ with $\mathcal{E}=0.25$. The green and red rectangles indicate the size of the uncertainties of the coherent states in valleys K (advancing on a solid blue spiral) and K' (dashed purple spiral), respectively. At the times $t = 15, 30$ and $45$, the coherent states advance differently on the spiral. The rectangle height and width indicate the uncertainty of $X$ and $P$ (conveniently, we decrease these rectangle lengths by the factor of 1/3). The fifth green rectangle corresponds to the quasi-period $t=107$ (see Eq. \eqref{period}). A movie shows the time evolution of the coherent states in \href{https://www.fis.unam.mx/~stegmann/WignerFunctions/Uncert.mp4}{\texttt{Uncert.mp4}}.}
\end{figure}

\begin{figure*}[t!!]
  \begin{tabular}{cc}
  (a) \qquad \qquad \qquad \qquad \qquad \qquad \qquad \qquad \qquad \qquad \qquad \qquad \qquad  & (b) \qquad \qquad \qquad \qquad \qquad \qquad \qquad \qquad \qquad \qquad \qquad \qquad \qquad \qquad \\
 \includegraphics[width=0.40\textwidth]{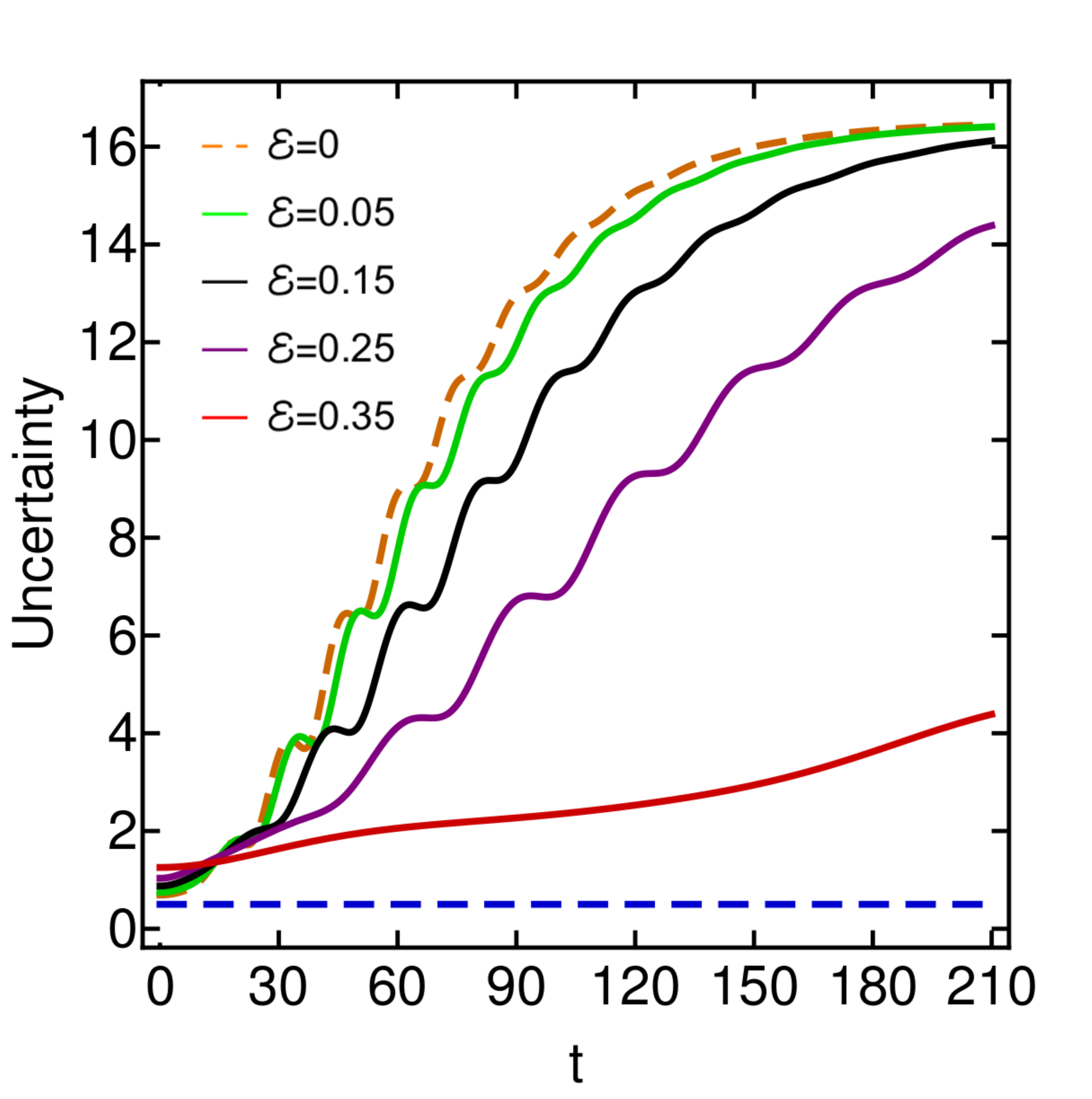} &
  \includegraphics[width=0.40\textwidth]{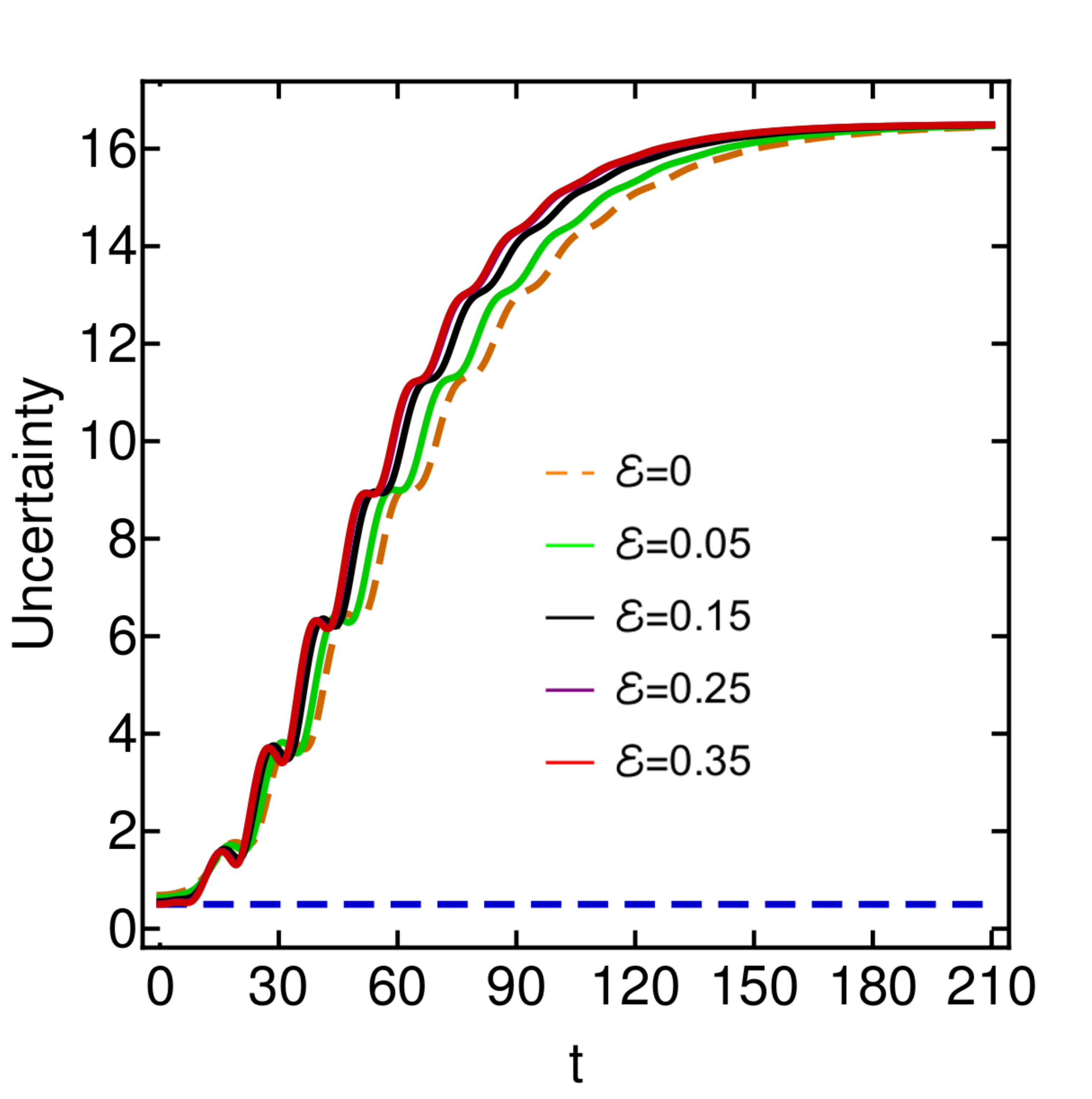}
  \end{tabular}
  \caption{Uncertainty relation $\Delta X \Delta P/\vert\langle[X,P]\rangle\vert$ as a function of the time $t$ in (a) valley K and (b) valley K' for different electric fields. In valley K', the uncertainty increases much faster as in valley K, where the uncertainty stays for a long time close to the minimal value of a coherent light state (blue dashed curve). All the uncertainties reach the same asymptotic value close to $|\alpha|^2$.}
  \label{Uncert}
\end{figure*}
\begin{figure}[ht]
	\centering
	\includegraphics[width=8cm]{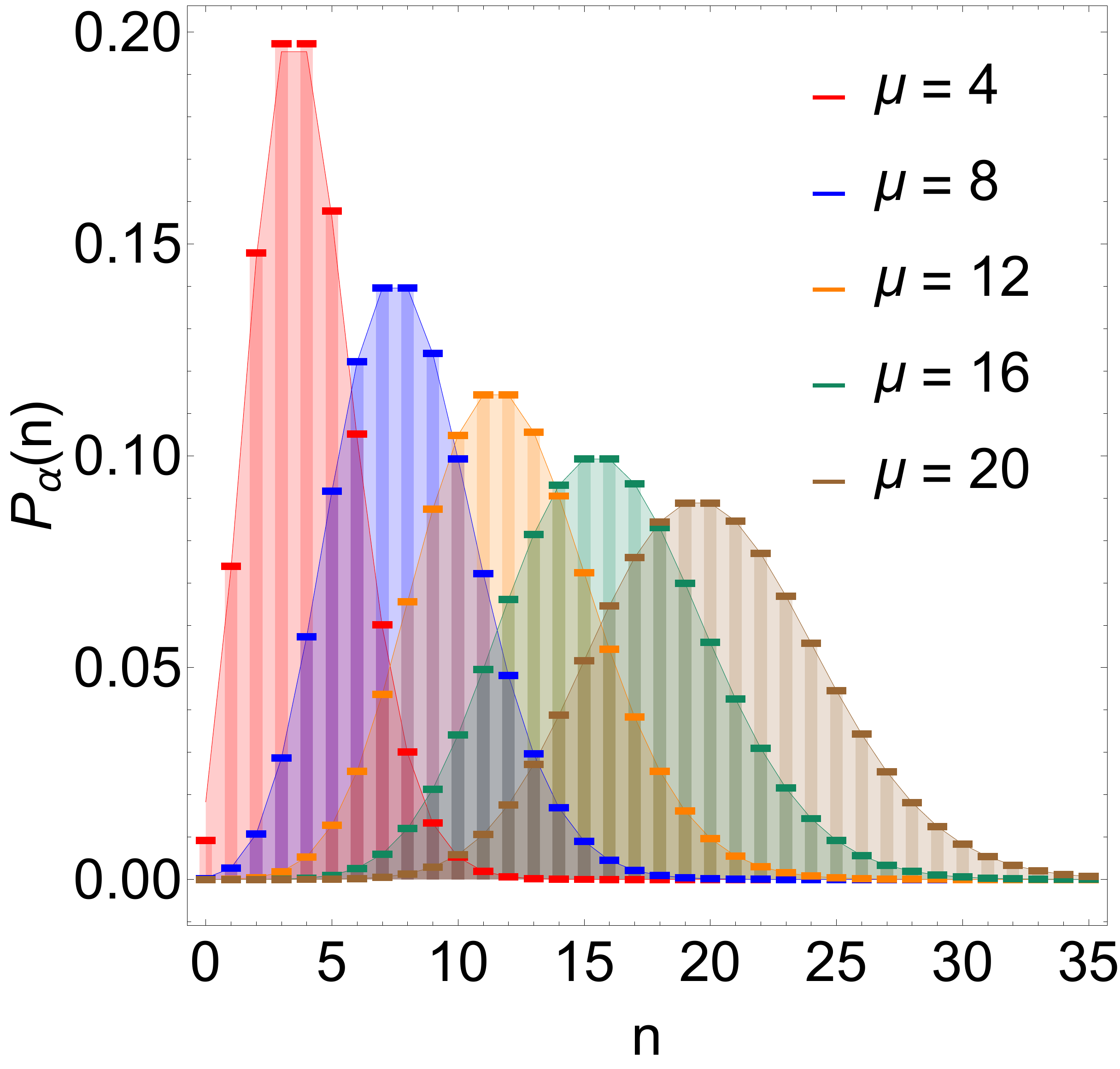}
	\caption{\label{fig:OND} Occupation number distribution $P_{\alpha}(n)$ in Eq.~\eqref{OND} for the coherent electron states $\Psi_{\alpha}$ for different values of $\mu=\vert\alpha\vert^{2}$. The solid lines connecting the dots, which show a Poisson distribution, are only guides to the eye and do not indicate continuity.}
\end{figure}
\begin{figure*}[p!!]
\begin{tabular}{c}
\includegraphics[trim = 0mm 0mm 0mm 0mm, scale= 0.21, clip]{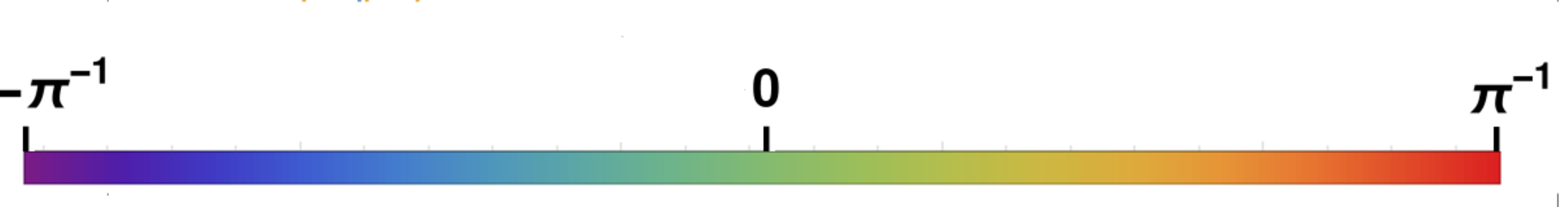}
\end{tabular}
\begin{tabular}{ccc}
(a) \qquad \qquad \qquad \qquad $t=0$, $\nu=1$ \qquad \qquad \qquad \qquad & \qquad \qquad \qquad &(b)\qquad \qquad \qquad \qquad $t=0$, $\nu=-1$ \qquad \qquad \qquad \qquad\\
\includegraphics[trim = 0mm 0mm 0mm 0mm, scale= 0.20, clip]{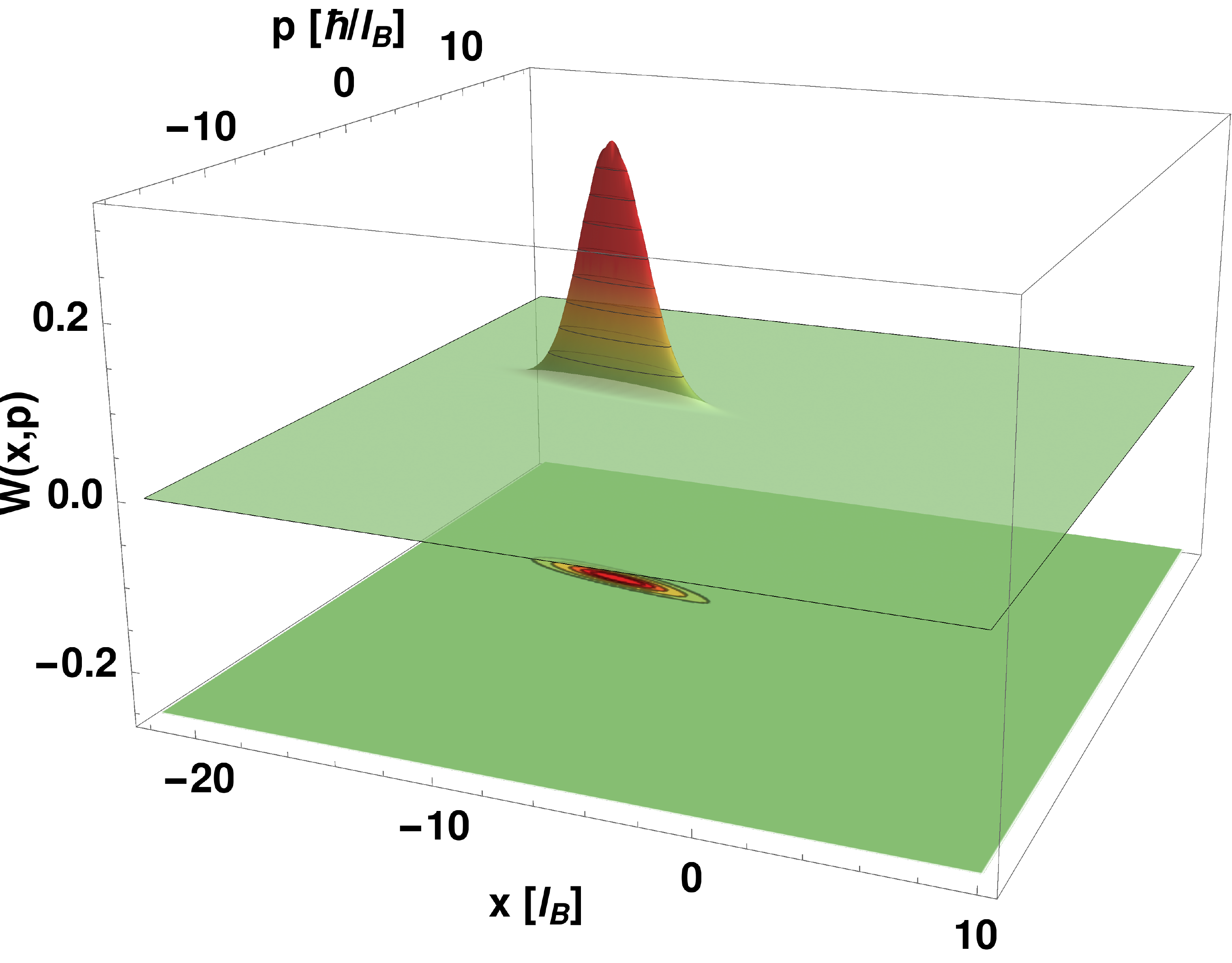} & \qquad \qquad \qquad &
\includegraphics[trim = 0mm 0mm 0mm 0mm, scale= 0.20, clip]{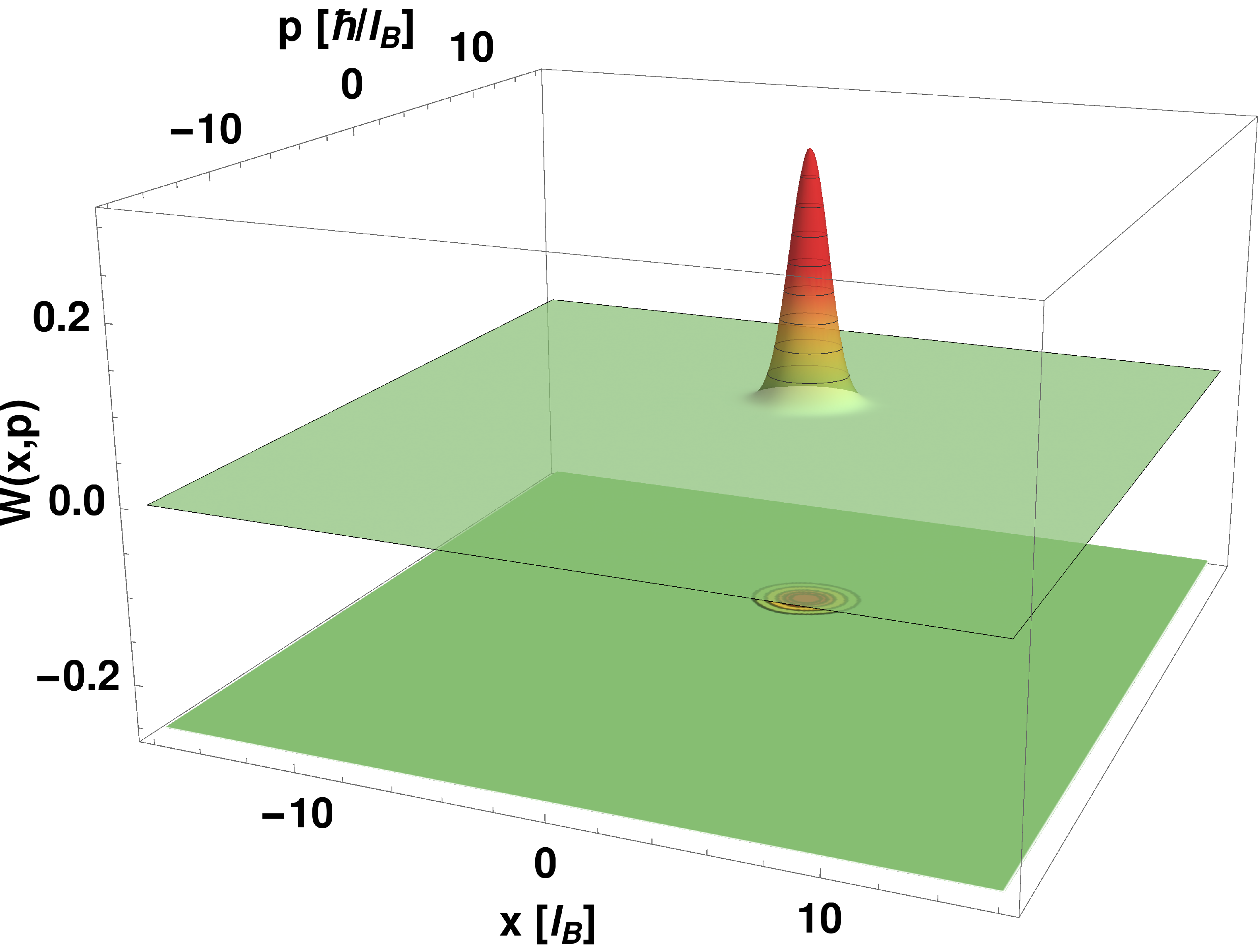}\\
(c) \qquad \qquad \qquad \qquad $t=15$, $\nu=1$ \qquad \qquad \qquad \qquad & \qquad \qquad \qquad & (d) \qquad \qquad \qquad \qquad $t=15$, $\nu=-1$ \qquad \qquad \qquad \qquad\\
\includegraphics[trim = 0mm 0mm 0mm 0mm, scale= 0.20, clip]{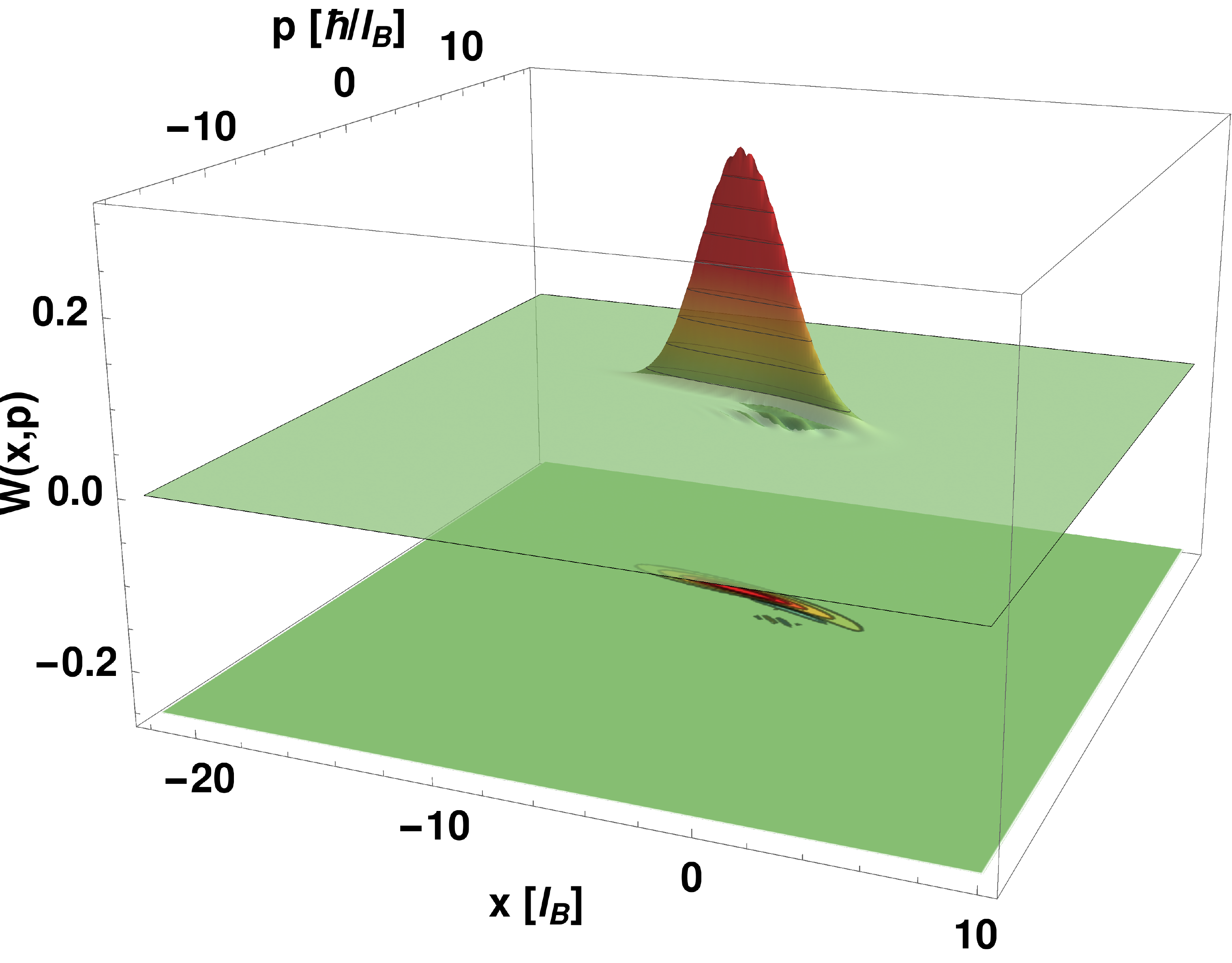} & \qquad \qquad \qquad &
\includegraphics[trim = 0mm 0mm 0mm 0mm, scale= 0.20, clip]{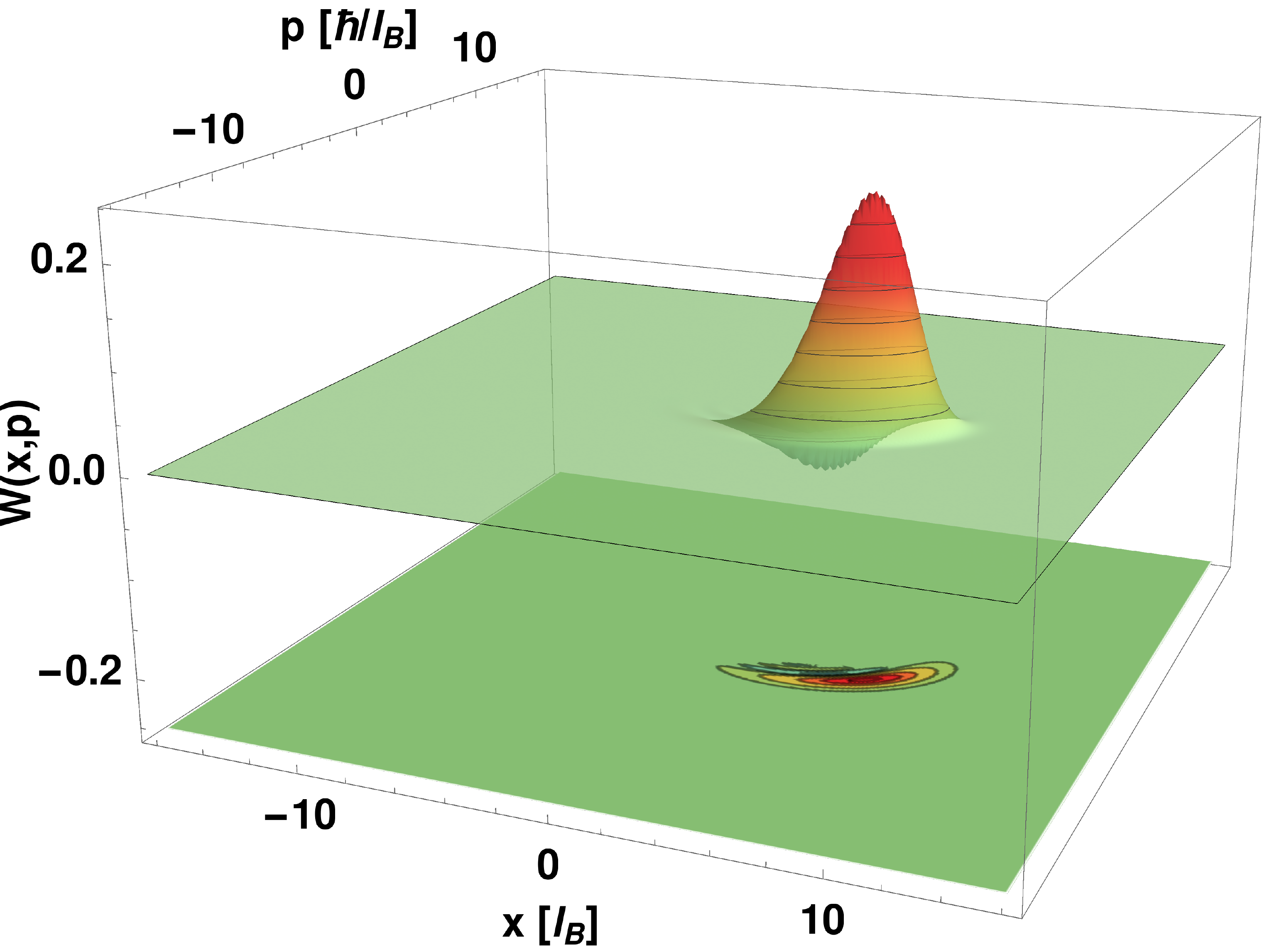}\\
(e) \qquad \qquad \qquad \qquad $t=30$, $\nu=1$ \qquad \qquad \qquad \qquad & \qquad \qquad \qquad & (f) \qquad \qquad \qquad \qquad $t=30$, $\nu=-1$ \qquad \qquad \qquad \qquad\\
\includegraphics[trim = 0mm 0mm 0mm 0mm, scale= 0.20, clip]{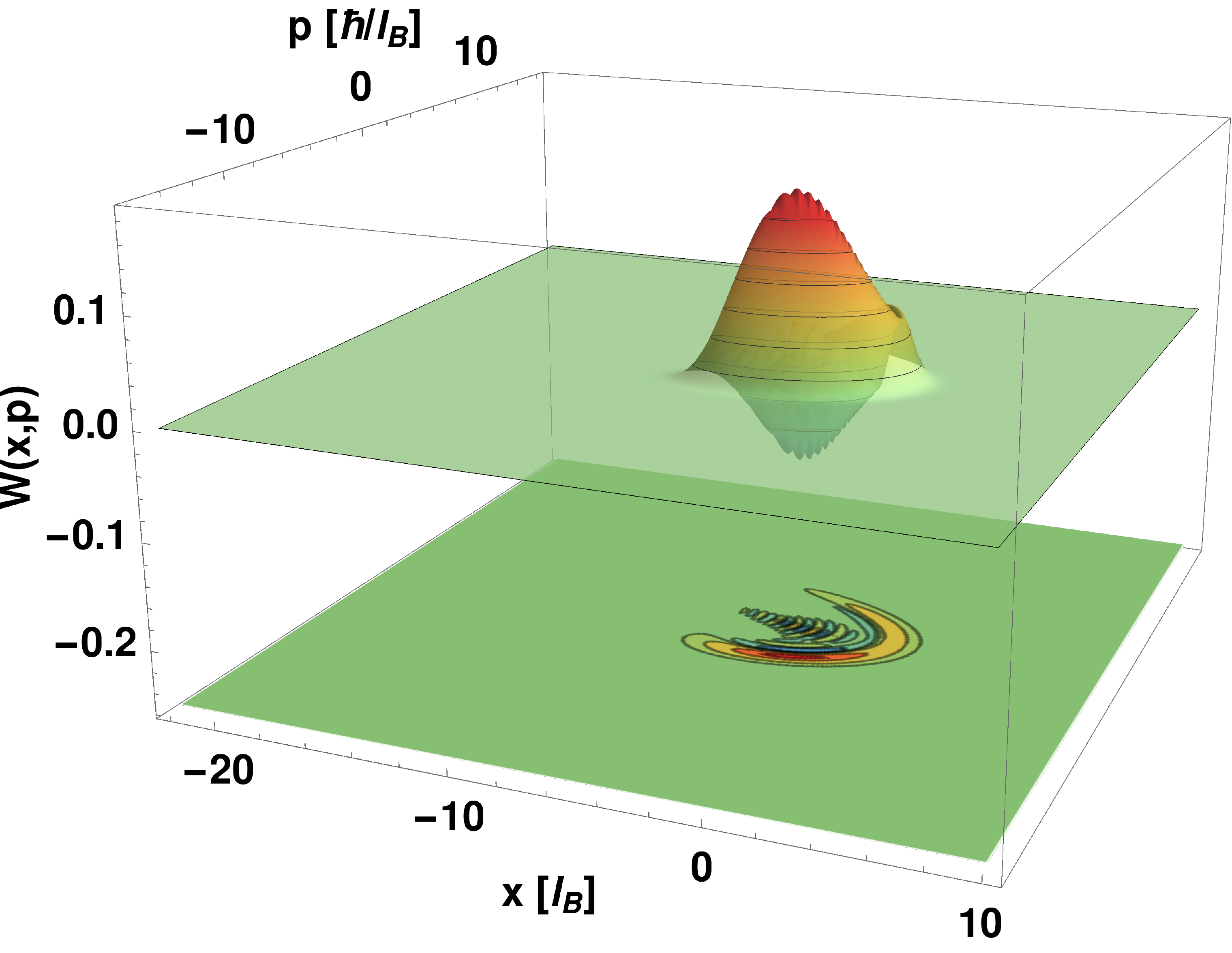} & \qquad \qquad \qquad &
\includegraphics[trim = 0mm 0mm 0mm 0mm, scale= 0.20, clip]{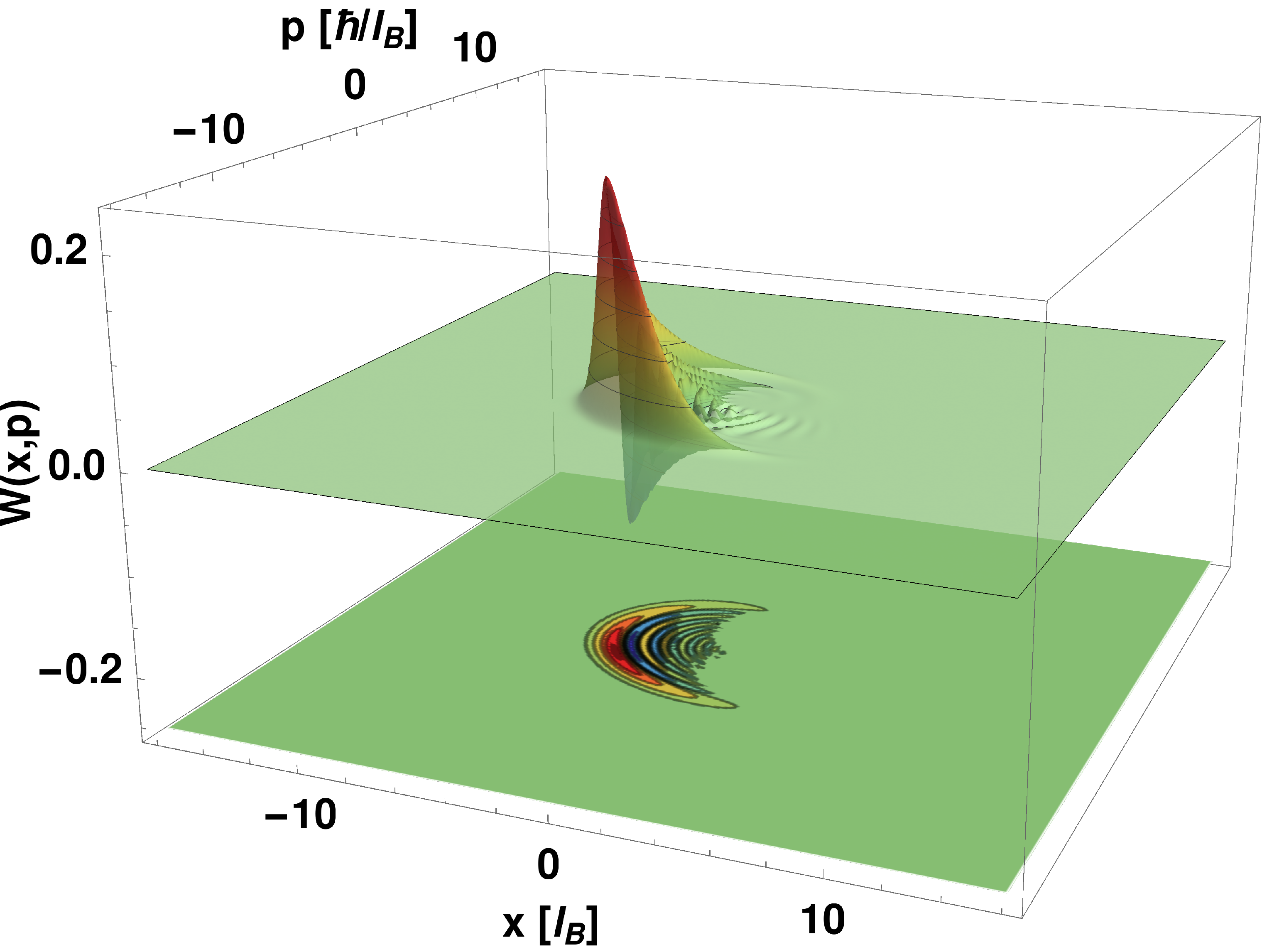}\\
(g) \qquad \qquad \qquad \qquad $t=45$, $\nu=1$ \qquad \qquad \qquad \qquad & \qquad \qquad \qquad & (h) \qquad \qquad \qquad \qquad $t=45$, $\nu=-1$ \qquad \qquad \qquad \qquad\\
\includegraphics[trim = 0mm 0mm 0mm 0mm, scale= 0.20, clip]{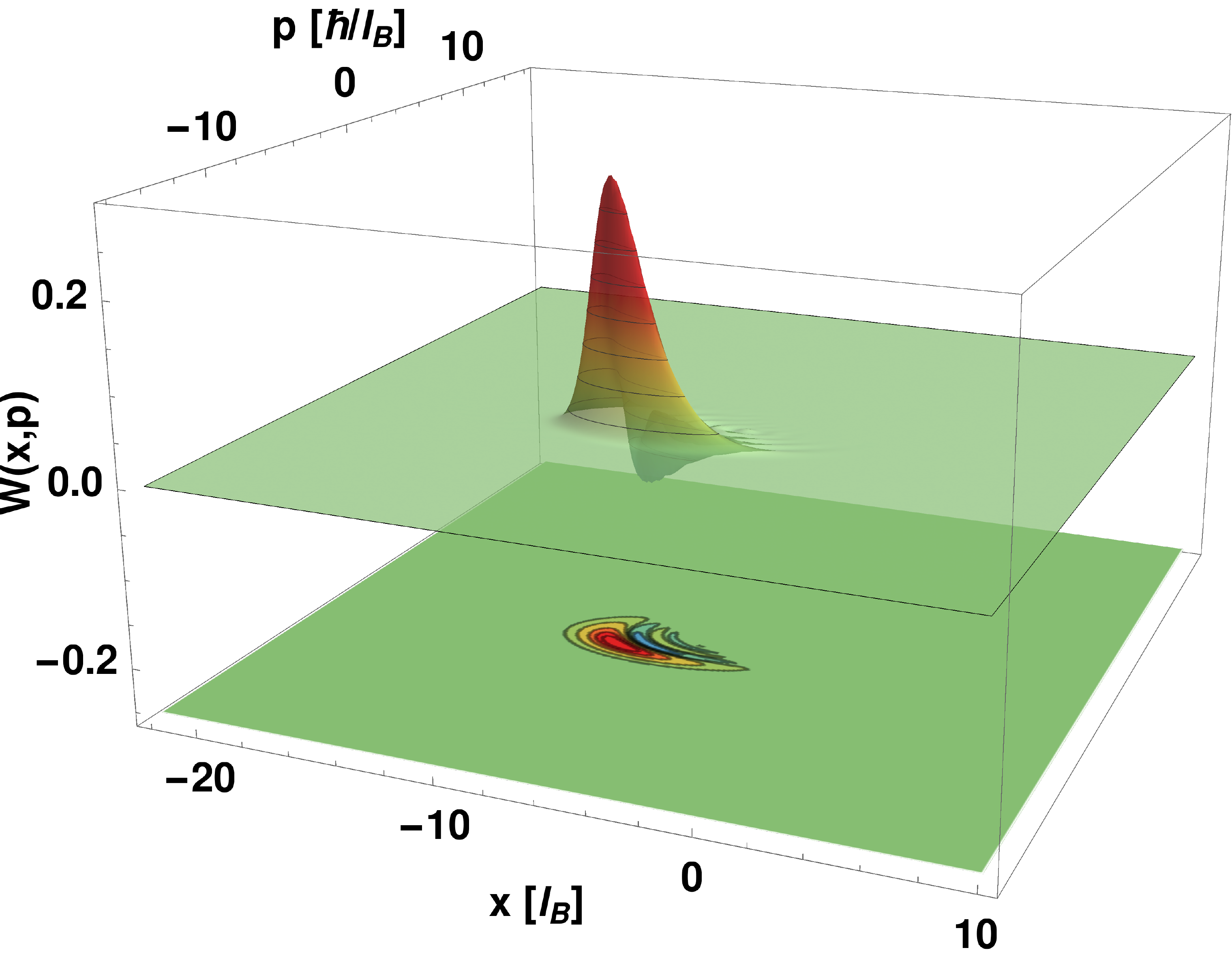} & \qquad \qquad \qquad &
\includegraphics[trim = 0mm 0mm 0mm 0mm, scale= 0.20, clip]{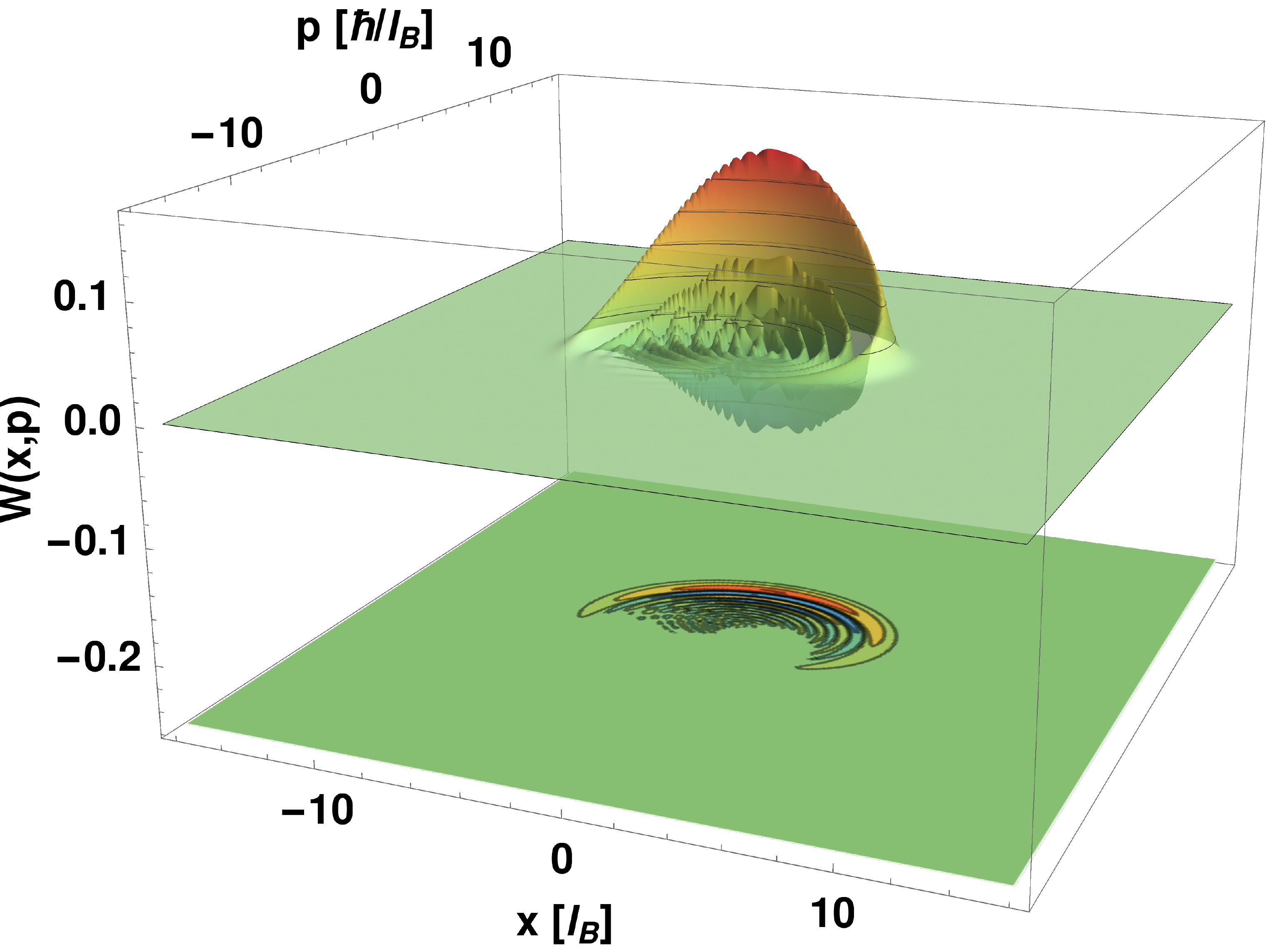}
\end{tabular}
\caption{Time evolution of the trace of the Wigner matrix $\mathbb{W}_{\alpha}(\vec{r},\vec{p})$ in Eq.~(\ref{traceWFt}) for different values of $t$ in each Dirac point ($\nu=\pm1$). $B=1$, $k_{y}=0$, $\alpha = 4i$, $v_x = 0.86$, $v_y = 0.69$, $v_t = 0.32$, $\mathcal{E}=0.25$, and $\lambda=1$. More details in the time evolution of the Wigner function are shown as movies in \href{https://www.fis.unam.mx/~stegmann/WignerFunctions/Wigner-nu+1.mp4}{\texttt{Wigner-nu+1.mp4}} and \href{https://www.fis.unam.mx/~stegmann/WignerFunctions/Wigner-nu-1.mp4}{\texttt{Wigner-nu-1.mp4}}.}
\label{TEWF}
\end{figure*}

\subsection{Time evolution of the Wigner function of coherent electron states}

In analogy with the standard harmonic oscillator, we define the position and momentum-like operators in terms of the creation and annihilation operators acting onto the Hilbert space basis \cite{Gerry2010}, which is expanded by the eigenstates of the Hamiltonian in Eq. \eqref{EMH}:
    \begin{equation}\label{operators}
    X  =  \frac{1}{\sqrt{2}}(\Theta^- + \Theta^+), \quad
    P  =  \frac{1}{\sqrt{2}i}(\Theta^- - \Theta^+). 
\end{equation}
\noindent We calculate from the time evolution of the coherent electron states in Eq. \eqref{timecoherentstate} the expectation values (see Appendix~\ref{App}):
\begin{subequations}\label{EQO}
  \begin{align}
    \langle X \rangle &= \frac{1}{\sqrt{2}(2\exp\left(\vert \alpha\vert^2\right)-1)}\left[C_{+}\Re(Z_{1})+\frac{C_{-}}{2}\Re(\mathcal{Z}_{1})\right], \\  \langle P \rangle &= \frac{1}{\sqrt{2}(2\exp\left(\vert \alpha\vert^2\right)-1)}\left[C_{+}\Im(Z_{1})+\frac{C_{-}}{2}\Im(\mathcal{Z}_{1})\right],
\end{align}   
\end{subequations}
where $\Im(z)$ denotes the imaginary part of a complex number $z$, and
\begin{subequations}
\begin{align}
Z_{m} & = \langle\Phi_{\alpha}\vert (\Theta^-)^m \vert\Phi_{\alpha}\rangle \nonumber \\
& = z^m\Bigg[2\sum^{\infty}_{n = 0} \frac{|\alpha|^{2n}}{n!}\textrm{e}^{-i(E_{n+m} - E_n)t}-\textrm{e}^{-i E_{m}t}\Bigg], \\
&& \nonumber\\
\mathcal{Z}_{m} & = \langle\Phi_{\alpha}\vert \sigma_{y}(\Theta^-)^m\sigma_{y} \vert\Phi_{\alpha}\rangle \nonumber \\
& = z^m\Bigg[\sum^{\infty}_{n = 0} \frac{|\alpha|^{2n}}{n!}\textrm{e}^{-i(E_{n+m} - E_n)t} \nonumber \\
& \qquad \qquad \times \left(\sqrt{\frac{n+m+1}{n+1}}+\sqrt{\frac{n}{n+m}}\right)\Bigg],
\end{align}
\end{subequations}
for $m = 1, 2, \ldots$. With the calculation of the expectation values of $\langle X \rangle$ and $\langle P \rangle$, we show the mean path of electrons in the presence of the crossed electric and magnetic fields in Fig.~\ref{fig:spiral}. It is worth noting that such a spiral trajectory depends on the valley index, which also affects the timescale. To estimate the generalized uncertainty principle (GUP),
\begin{equation}
    \Delta X\,\Delta P\geq\frac{\vert\langle[X,P]\rangle\vert}{2},
\end{equation}
we have also computed the following expressions:
\begin{subequations}
    \begin{align}
&(\Delta X)^{2}  =  \langle X^2 \rangle-\langle X \rangle^2 \nonumber \\
 & =  \frac{\mathcal{N}_{\alpha}^{2}}{4}\bigg\{C_{+}\left[4(\vert z\vert^{2}+1)\textrm{e}^{\vert\alpha\vert^{2}}-\vert z\vert^{2}-3+2\Re\left(Z_{2}\right)\right] \nonumber \\
&\qquad +C_{-}\left[(\vert z\vert^{2}+1)\textrm{e}^{\vert\alpha\vert^{2}}+\frac12\Re\left(\mathcal{Z}_{2}\right)\right]\bigg\} - \langle X \rangle^2, \\
&(\Delta P)^{2} =  \langle P^2 \rangle-\langle P \rangle^2 \nonumber\\
 & =  \frac{\mathcal{N}_{\alpha}^{2}}{4}\bigg\{C_{+}\left[4(\vert z\vert^{2}+1)\textrm{e}^{\vert\alpha\vert^{2}}-\vert z\vert^{2}-3-2\Re\left(Z_{2}\right)\right] \nonumber \\
&\qquad +C_{-}\left[(\vert z\vert^{2}+1)\textrm{e}^{\vert\alpha\vert^{2}}-\frac12\Re\left(\mathcal{Z}_{2}\right)\right]\bigg\} - \langle P \rangle^2, \\
&\mathcal{N}_{\alpha}^{2}=\frac{1}{2\exp\left(\vert \alpha\vert^2\right)-1},
\end{align}
\end{subequations}
\noindent which are represented by the rectangles on the spiral in Fig.~\ref{fig:spiral} for the times $t = 15, 30,$ and $45$. The states at the two valleys advance differently on the spiral, and the uncertainties increase towards the spiral center. During the time evolution, the uncertainty relation $\Delta X \Delta P/\vert\langle[X,P]\rangle\vert$ does not keep the minimum value, as shown in Fig. \ref{Uncert}. The uncertainty increases in both valleys, though at a much slower rate in valley K, where the uncertainty stays for longer times close to the minimal value of a coherent light state (see the dashed blue horizontal line). This behavior can be explained by the fact that the relative separation between Landau levels is almost constant for any index level $n$, which approximates the constant separation between energy levels in the quantum harmonic oscillator. The maximum value of the uncertainty is approximately given by $|\alpha|^2$, where $\vert\alpha\vert$ gives the amplitude of the oscillations~\cite{Cohen}. Besides, we can verify from the probability distribution of occupation number $P_{\alpha}(n)=\vert\langle\Psi_{n}\vert\Psi_{\alpha}\rangle\vert^{2}$ that the coherent electron states follow a Poisson-like distribution with mean $\mu=\vert\alpha\vert^{2}$ (see Fig.~\ref{fig:OND}):
\begin{equation}\label{OND}
    P_{\alpha}(n)=\frac{1}{2\exp\left(\mu\right)-1}
    \begin{cases}
    1, & n=0, \\
    \frac{2\mu^{n}}{n!}, & n>0.
    \end{cases}
\end{equation}
\noindent The maximum uncertainty occurs when the prepared coherent state is getting closer to the center of the spiral in Fig.~\ref{fig:spiral}. This coherent state tends to look like the Landau one that contributes the most to the linear combination in (\ref{40}), according to the distribution of occupation numbers $P_{\alpha}(n)$, where $n$ is equal to the integer part of $\mu=\vert\alpha\vert^{2}$. These features also appear in the time evolution of the WF of coherent states, as will be explained later on.

Furthermore, the WF is calculated using the components of coherent states in Eq. \eqref{timecoherentstate} into the integral matrix representation in Eq. \eqref{WM}:
\begin{equation}
\mathbb{W}_{\alpha}(\vec{r},\vec{p})=\mathbb{M}W_{\alpha}(\vec{r},\vec{p})\mathbb{M}^{\dagger}.
\end{equation}
Thus, the trace of this matrix provides us an expression of the time-dependent WF for coherent states:
\begin{eqnarray}\label{traceWFt}
\nonumber{\rm Tr}[\mathbb{W}_{\alpha}(\vec{r},\vec{p},t)]&=&\frac{\delta\left(p_{y}-k_{y}\right)}{2\exp\left(\vert \alpha\vert^2\right)-1}\Big\{W_{11}(\chi,t) \nonumber\\
&& \qquad +W_{22}(\chi,t)-2\lambda\beta_{\nu}\Re[W_{12}(\chi,t)]\Big\}, \nonumber\\
&&
\end{eqnarray}
where 
\begin{subequations}\label{componentst}
	\begin{align}
	W_{11}(\chi,t)& =\sum_{n,m=1}^{\infty}\frac{\alpha^{n}\alpha^{\ast m}\textrm{e}^{i(E_m-E_n)t}}{\sqrt{n!m!}}W_{n-1,m-1}(\chi_{n,m}),\\
	W_{12}(\chi,t)&=\sum_{n=1}^{\infty}\sum_{m=0}^{\infty}\frac{\alpha^{n}\alpha^{\ast m}\textrm{e}^{i(E_m-E_n)t}}{\sqrt{n!m!}}W_{n-1,m}(\chi_{n,m}) \nonumber \\
	&=W^\ast_{21}(\chi,t),\\
	W_{22}(\chi,t)&=\sum_{n,m=0}^{\infty}\frac{\alpha^{n}\alpha^{\ast m}\textrm{e}^{i(E_m-E_n)t}}{\sqrt{n!m!}}W_{n,m}(\chi_{n,m}).
	\end{align}
\end{subequations}

The time evolution of the WF for these states in Fig. \ref{TEWF} agrees with the spiral behavior of the expectation value of the operators $X$ and $P$ in Fig.~\ref{fig:spiral}.  We observe that the WF in valley K propagates slower by the factor $(1 - \beta^2_\nu)^{3/4}$ than in valley K', preserving a positive value for a longer time. If this factor tends to zero, the coherent electron state stays fixed at the same point in phase space and presents a Gaussian distribution. This is in contrast to coherent light states, where the minimum uncertainty remains indefinitely in a circular trajectory, and all the values of the WF are positive. Negative values of the WF appear when the state advances toward the origin of the spiral. This behavior is related closely to the increase of the uncertainty relations to the asymptotic value $\sim |\alpha|^2$, where the occupation probability has a Poisson distribution (see Fig.~\ref{fig:OND}). Nevertheless, there are also evident differences. For instance, we can note that the initial states in Figs.~\ref{fig:spiral} and \ref{TEWF} have a different position in phase space. For larger times, as shown in \href{https://www.fis.unam.mx/~stegmann/WignerFunctions/Wigner-nu+1.mp4}{\texttt{Wigner-nu+1.mp4}} and \href{https://www.fis.unam.mx/~stegmann/WignerFunctions/Wigner-nu-1.mp4}{\texttt{Wigner-nu-1.mp4}}, the WFs are identical to that of the Landau state with $n$ equal to the integer part of $|\alpha|^2$ for valleys K and K', respectively, turning around the centers given by $\sqrt{2}\beta_\nu|\alpha|$ (see Eq.~\eqref{xi}). This singular feature, which also appears for the WF of Landau states in Fig.~\ref{fig2}, is a distinctive signature of the tilting of Dirac cones. Without electric field, the centers are located at the symmetrical points $\nu\sqrt{2}v_t|\alpha|/v_y$. The drift velocity $\vec{v}_{\rm d}$ separates asymmetrically WFs in phase space. Such a signature remains hidden in Fig.~\ref{fig:spiral}, showing an identical center due to the definition of $X$ and $P$ as scaled and shifted operators in Eqs.~\eqref{operators}.

Now, to find a possible approximate period $\tau$ for coherent states, we can proceed as in Ref.~\cite{Fernandez2020}. Setting the eigenvalue $\alpha$, we compute the mean energy value $\langle H\rangle_{\alpha}$ and the energy interval in which it lies, namely, $E_{j,k_{y}}<\langle H\rangle_{\alpha}<E_{j+1,k_{y}}$. Thus, the approximate period is determined as:
\begin{equation}\label{defperiod}
\tau=\frac{2\pi}{E_{j+1,k_{y}}-E_{j,k_{y}}}.
\end{equation}
As the energy spectrum depends on tilting parameter $\nu$ through parameter $\beta_{\nu}$, the period $\tau$ will be different for each valley.

The mean energy $\langle H\rangle_{\alpha}$ for the coherent states $\Psi_{\alpha}(x,y)$ is given by
\begin{align}\label{energy1}
\nonumber&\langle H\rangle_{\alpha}=\frac{1}{2\exp\left(\vert \alpha\vert^2\right)-1}\Bigg[k_{y}\frac{\mathcal{E}}{B}\left(1-2\exp\left(\vert\alpha\vert^2\right)\right)\nonumber\\
& \quad +\text{sgn}(n)\frac{2\sqrt{v_{x}v_{y}}(1-\beta_{\nu}^2)^{3/4}}{l_{\rm B}}\sum_{n=1}^{\infty}\frac{\vert\alpha\vert^{2n}}{n!}\sqrt{2|n|}\,\Bigg].
\end{align}
For instance, for the coherent states with eigenvalue $\alpha=4i$, and the same values used in Figs.~\ref{fig2} and~\ref{TEWF}, we have $E_{15}<\langle H\rangle=1.828<E_{16}$ for $\nu=+1$ and $E_{15}<\langle H\rangle=4.289<E_{16}$ for $\nu=-1$. Therefore, the respective quasi-periods turn out to be:
\begin{equation}\label{period}
\tau_{+}\approx34.165\pi, \quad \tau_{-}\approx14.566\pi.
\end{equation}
These results agree with the behavior shown in Figs.~\ref{fig:spiral} and \ref{TEWF}, in which the times $t=107$ and $t=45$ have been considered for the coherent states at valleys K and K', respectively. The time dilation on one of the valleys for coherent states is shown as movies in the Supplemental Material. In general, the period $\tau$ in~(\ref{defperiod}) increases as the separation of the energy levels in~(\ref{energy}) decreases close to the critical value $\mathcal{E}_{c}$ of the electric field. This can be attributed to the longer time it takes a Dirac fermion to complete a loop, as its orbit tends to open. While for more separated energy levels, far away from the critical value $\mathcal{E}_{c}$, the period $\tau$ is less, indicating that orbits are closed.

\section{Conclusions and final remarks}
Anisotropic and tilted Dirac cone materials possess evident valley-dependent electronic properties under crossed electric and magnetic fields. A simplified continuum model captures the main features of the electronic band structure of several two-dimensional materials with anisotropic and tilted Dirac cones. The effective Hamiltonian depends on two anisotropic velocities and one tilt velocity (see Eq. \eqref{TAH}). In particular, we set these parameters to the known values of the 8-$pmmn$ borophene in all our calculations. We use a canonical transformation for mapping the anisotropic and tilted Dirac Hamiltonian under crossed fields to an isotropic one without the tilting of Dirac cones (see Eqs. \eqref{EMH}, \eqref{CT}, and \eqref{Hc}). This allowed using the well-known solutions of the isotropic case in our current configuration with anisotropy and Dirac cone tilt (see Eqs. \eqref{energy} and \eqref{LS}). We obtained Landau levels and their states, which are generally not degenerated for valleys K and K'. We showed that the WF for the Landau state $n = 0$ has a Gaussian shape with a different deformation according to the valley index (see Figs. \ref{fig2}(a) and (b)). In all Landau states, the valley-dependent factor $\beta_\nu$ modulates the deformation of the WFs through the electric and magnetic fields. The valley degeneracy of Landau levels in the absence of an electric field avoids identifying the valley dependence in density of states and magnetoconductivity in tilted Dirac materials \cite{Islam2018}. An important advantage for the realization of quantum tomography experiments is the possibility to identify the valley dependence of WFs still without the presence of an electric field.

We developed the coherent states using the Landau eigenfunction basis of the Hamiltonian in Eq. \eqref{EMH}. We also proposed the position and momentumlike operators in terms of the annihilation operator in Eqs. \eqref{QO}. We evidenced that the expectation values of the operators $X$ and $P$ in  phase space and the time evolution of the WF follow a spiral behavior, as shown in Figs.~\ref{fig:spiral} and \ref{TEWF}. The emergence of negative values in the WF is related to the increasing uncertainties of the position and momentum. The coherent electron states lose the coherence, and tend to the Landau state that contributes the most to the linear combination of the coherent state itself, Eq. \eqref{40}. The mean values of $X$ and $P$ go towards the origin of the spiral, as shown in Fig. \ref{fig:spiral} and Supplemental Material \href{https://www.fis.unam.mx/~stegmann/WignerFunctions/Uncert.mp4}{\texttt{Uncert.mp4}}, and it is due to the definition of these operators in Eqs. \eqref{operators}. However, the time evolution of WFs in Fig. \ref{TEWF} and Supplemental Material (\href{https://www.fis.unam.mx/~stegmann/WignerFunctions/Wigner-nu+1.mp4}{\texttt{Wigner-nu+1.mp4}} and \href{https://www.fis.unam.mx/~stegmann/WignerFunctions/Wigner-nu-1.mp4}{\texttt{Wigner-nu-1.mp4}}) showed that coherent states reached different positions in phase space at longer times, which can be tested from quantum tomography experiments. Increasing the electric field to a critical value causes a time dilation in the WF in one of the valleys (see Figs.~\ref{fig:spiral}, \ref{Uncert}, and \ref{TEWF}). We also estimated a quasiperiod in the time evolution of coherent states from the expectation value of the Hamiltonian in Eqs. \eqref{period}. Thus, critical values of the electric field do not only collapse the WFs and the Landau-level spectra in a single valley but also delay the time evolution of coherent states in a small region of the phase space, keeping a minimum uncertainty. It is also worth remarking that, although our numerical results were computed by using the 8-$pmmn$ borophene parameters, they can be extended to any other tilted anisotropic Dirac cone material, by adjusting the corresponding parameters. Our theoretical findings may help to establish experimental protocols for the realization of coherent electron states in two-dimensional materials under the interaction of electromagnetic fields. For instance, the coherent state description developed through the phase-space representation may provide a way of describing phenomena such as the Shubnikov-de Hass oscillations via quantum tomography experiments in the presence of a Hall field at low temperatures.

\section*{Acknowledgments}

We acknowledge financial support from CONACYT Project No. A1-S-13469, CONACYT Project No. FORDECYT-PRONACES/61533/2020, the UNAM-PAPIIT Research Grant IA 103020, and the SIP-IPN Research Grant 20210317.
 
 \appendix
 
\section{Generalized uncertainty principle}\label{App}
To compute the uncertainties $\Delta X$ and $\Delta P$ of the operators $X$ and $P$ in Eqs.~(\ref{operators}) and to establish the generalized uncertainty principle,
\begin{equation}\label{GUP}
    \Delta X\,\Delta P\geq\frac{\vert\langle[X,P]\rangle\vert}{2},
\end{equation}
we define the following operators and its square:
\begin{subequations}\label{A1}
    \begin{align}
    \Upsilon_{q}&=\frac{1}{\sqrt{2}i^{q}}(\Theta^{-}+(-1)^{q}\Theta^{+}),\\
    \Upsilon_{q}^{2}&=\frac12\left[\Theta^{-}\Theta^{+}+\Theta^{+}\Theta^{-}+(-1)^{q}\left((\Theta^{-})^{2}+(\Theta^{+})^{2})\right)\right],
\end{align}
\end{subequations}
for $q=0,1$, such that $\Upsilon_{0}\equiv X$ and $\Upsilon_{1}\equiv P$.

From Eq.~(\ref{31}), we have that $\Psi_{\alpha}(x,y)=\mathbb{M}\Phi_{\alpha}(x,y)$, where
\begin{equation}
    \mathbb{M}=\sqrt{\frac{1}{2}}\left(\sqrt{C_{+}}\sigma_{0}-\sqrt{C_{-}}\sigma_{y}\right).
\end{equation}
Thus,
\begin{align}\label{A2}
\nonumber    \langle\Psi_{\alpha}\vert\Upsilon_{q}\vert\Psi_{\alpha}\rangle&=\langle\Phi_{\alpha}\vert\mathbb{M}^{\dagger}\Upsilon_{q}\mathbb{M}\vert\Phi_{\alpha}\rangle\\
\nonumber    &=\frac12\Big[C_{+}\langle\Phi_{\alpha}\vert\Upsilon_{q}\vert\Phi_{\alpha}\rangle+C_{-}\langle\Phi_{\alpha}\vert\sigma_{y}\Upsilon_{q}\sigma_{y}\vert\Phi_{\alpha}\rangle\\
&\quad -\sqrt{C_{+}C_{-}}\langle\Phi_{\alpha}\vert(\Upsilon_{q}\sigma_{y}+\sigma_{y}\Upsilon_{q})\vert\Phi_{\alpha}\rangle\Big].
\end{align}

Using the simplest notation $\vert n;k\rangle\equiv\vert\psi_{n}(\xi_{k})\rangle$ and taking into account that the operators $\mathcal{T}^{\pm}$ might generate states not physically acceptable, as well as the orthogonality condition $\langle n';k'\vert n;k\rangle=\delta_{n'n}\delta_{k'k}$, the mean value of $\Upsilon_{q}$ and its square $\Upsilon_{q}^{2}$ turn out to be
\begin{subequations}
\begin{align}
    \langle\Phi_{\alpha}\vert\Upsilon_{q}\vert\Psi_{\alpha}\rangle&=\frac{\mathcal{N}_{\alpha}^{2}}{2\sqrt{2}i^{q}}\bigg[C_{+}\left(f(z,t)+(-1)^{q}f^{\ast}(z,t)\right) \nonumber\\
    &\quad+\frac{C_{-}}{2}\left(g(z,t)+(-1)^{q}g^{\ast}(z,t)\right)\bigg],\\
  \langle\Phi_{\alpha}\vert\Upsilon_{q}^{2}\vert\Psi_{\alpha}\rangle&=\frac{\mathcal{N}_{\alpha}^{2}}{4}\bigg[C_{+}\Big(4(\vert z\vert^{2}+1)\textrm{e}^{\vert\alpha\vert^{2}}-\vert z\vert^{2} \nonumber\\
  &\quad-3+2(-1)^{q}\Re\left(r(z,t)\right)\Big) \nonumber\\
  &\quad+C_{-}\left((\vert z\vert^{2}+1)\textrm{e}^{\vert\alpha\vert^{2}}\right. \nonumber\\
  &\qquad+\left.\frac{(-1)^{q}}{2}\Re\left(s(z,t)\right)\right)\bigg],
\end{align}
\end{subequations}
where
\begin{subequations}
    \begin{align}
        f(z,t)&=z\left[2\sum_{n=0}^{\infty}\frac{\vert\alpha\vert^{2n}\textrm{e}^{-i(E_{n+1}-E_{n})t}}{n!}-\textrm{e}^{-iE_{1}t}\right], \\
        g(z,t)&=z\sum_{n=0}^{\infty}\frac{\vert\alpha\vert^{2n}\textrm{e}^{-i(E_{n+1}-E_{n})t}}{\sqrt{n!(n+1)!}}(\sqrt{n+2}+\sqrt{n}), \\
        r(z,t)&=z^{2}\left[2\sum_{n=0}^{\infty}\frac{\vert\alpha\vert^{2n}\textrm{e}^{-i(E_{n+2}-E_{n})t}}{n!}-\textrm{e}^{-iE_{2}t}\right], \\
        \nonumber s(z,t)&=z^{2}\sum_{n=0}^{\infty}\frac{\vert\alpha\vert^{2n}\textrm{e}^{-i(E_{n+2}-E_{n})t}}{\sqrt{n!(n+2)!}}\\
        &\quad\times(\sqrt{(n+2)(n+3)}+\sqrt{n(n+1)}).
    \end{align}
\end{subequations}

Finally, the GUP reads
\begin{equation}
    \Delta X\,\Delta P\geq\frac{\mathcal{N}_{\alpha}^{2}}{4}\left\vert C_{+}\left[4\textrm{e}^{\vert\alpha\vert^{2}}+\vert z\vert^{2}-3\right]+C_{-}\textrm{e}^{\vert\alpha\vert^{2}}\right\vert.
\end{equation}

\end{document}